\documentclass[prb,reprint,twocolumn,amsmath,amssymb]{revtex4-1}

\usepackage{graphicx}
\usepackage{dcolumn}
\usepackage{bm}

\DeclareMathOperator{\R}{Re}

\begin{document}

\preprint{APS/123-QED}

\title{$\mathcal{PT}$-Symmetry Breaking and Catastrophes in Dissipationless Resonant Tunneling Heterostructures}

\author{A.\,A.\,Gorbatsevich}
 \affiliation{P.N. Lebedev Physical Institute of the Russian Academy of Sciences, 119991, Moscow, Russia.}
 \altaffiliation[Also at ]{National Research University of Electronic Technology, 124498, Zelenograd, Moscow, Russia.}
 \email{aagor137@mail.ru}

\author{N.\,M.\,Shubin}%
\affiliation{National Research University of Electronic Technology, 124498, Zelenograd, Moscow, Russia.}

\date{\today}

\begin{abstract}

We study the phenomenon of spontaneous symmetry breaking in dissipationless resonant tunneling heterostructures (RTS). To describe the quantum transport in this system we apply both the nonequilibrium Green function formalism based on a tight-binding model and a numerical solution of the Schr\"odinger equation within the envelope wavefunction formalism. An auxiliary non-Hermitian Hamiltonian is introduced. Its eigenvalues determine exactly the transparency peak positions. We present a procedure how to construct a family of non-Hermitian Hamiltonians with real eigenvalues. In general these Hamiltonians do not have $\mathcal{PT}$-symmetry. In spatially symmetric RTS the corresponding auxiliary non-Hermitian Hamiltonian becomes $\mathcal{PT}$-symmetric and possesses real eigenvalues, which can coalesce at exceptional points (EP) of Hamiltonian. A coalescence of the auxiliary non-Hermitian Hamiltonian eigenvalues means a coalescence of resonances in RTS, which is accompanied be symmetry breaking of the electron wavefunction probability distribution (at a given direction of the particle flow). We construct a classification of different types of the peak coalescence in terms of the catastrophe theory and investigate the impact of dissipation and asymmetry on these phenomena. Possible applications include sensors and broad-band filters.
\end{abstract}

\pacs{Valid PACS appear here}
\maketitle

\section{Introduction}
A new class of symmetry breaking (SB) phenomena has attracted a lot of attention in past years -- $\mathcal{PT}$-symmetry breaking\cite{bib:PTBend98,bib:PTBend99,bib:PTBend07} (here $\mathcal{P}$ stands for the space inversion and $\mathcal{T}$ -- for the time reversal symmetry operations). A new physical field has emerged following the discovery of Bender \emph{et al.}\cite{bib:PTBend98} that a non-Hermitian Hamiltonian, which is  invariant with respect to both the space inversion ($\mathcal{P}$) and the time reversal ($\mathcal{T}$) can possess real eigenvalues. These Hamiltonians are referred to as pseudo-Hermitian Hamiltonians\cite{bib:mostafa2002}. SBs in condensed matter physics are closely related to phase transitions. Microscopic mechanisms of phase transitions are quite cumbersome and involve as a rule elaborate many-body interactions. On the contrary, mechanisms of $\mathcal{PT}$-SB look very simple.  Under a variation of a specific system parameter (tuning parameter), which doesn't change directly the symmetry of the system, two real eigenvalues can coalesce and transform into other two ones with nonzero imaginary parts of different signs and with equal real parts, which is called $\mathcal{PT}$-symmetry breaking ($\mathcal{PT}$-SB). Such behavior of eigenvalues was known long ago in linear operators theory where the point in the parameter space, at which two real eigenvalues coalesce and turn into a pair of complex eigenvalues, is called exceptional points (EP)\cite{bib:BookKato,bib:Heiss,bib:Berry}. Besides their fundamental physical interest systems with EP in energy spectrum are claimed to be promising in different applications, such as sensors\cite{bib:EPWier}, non-reciprocal light transmission lines \cite{bib:NonRec, bib:PTopt2} and robust asymmetric waveguide switches \cite{bib:Switch}. 

Up to now the most realistic applications of $\mathcal{PT}$-SB with possible experimental manifestations have been based on the formal equivalence of the Schr\"oedinger and wave equations and described electromagnetic phenomena\cite{bib:Zyab}. A Hamiltonian eigenvalue with nonzero imaginary part corresponds to a nonunitary wavefunction evolution, which is forbidden for fermions by norm preserving condition. Hence, it  was unclear whether fermionic systems having anything to do with pseudo-Hermitian $\mathcal{PT}$-invariant Hamiltonians could exist. However, it is well known that in optics $\mathcal{T}$-breaking terms in the wave equation describe well established gain/loss processes. $\mathcal{PT}$-symmetric optical systems were studied with great detail, including waveguide\cite{bib:PTopt1, bib:PTopt2} and photonic lattice\cite{bib:NatPTLight} structures, laser systems\cite{bib:Laser, *bib:nature} and a new type of optically active systems -- coherent perfect absorbers\cite{bib:CPACho,*bib:CPALon,*bib:Cho}. Features of scattering problems in $\mathcal{PT}$-symmetric optical systems were studied in detail in Ref.\cite{bib:Ambi,*bib:PTmicroW,*bib:Feng2016}. Also $\mathcal{T}$-breaking terms as creation/annihilation processes can be easily introduced in bosonic systems. $\mathcal{PT}$-symmetric Bose-Einstein condensate was studied in Ref.\cite{bib:BEC}. Superconducting $\mathcal{PT}$-invariant model was considered in Ref.\cite{bib:PTsc}. But $\mathcal{T}$-breaking terms in this case describe creation/annihilation processes once again in a bosonic (Cooper-pair) field. Special case of exotic Majorana fermions was described in Ref.\cite{bib:Madrid,*bib:M44}.
 
An eigenstate of the Hamiltonian at EP is nondegenerate contrary to a crossing point. So, formally the number of states reduces from two to one under the variation of a tuning parameter. Such transformation is forbidden in a closed system that conserves particle number. Hence openness is a necessary condition for a physical realization of fermionic EP. Discrete energy levels of a closed quantum system match the resonances in the corresponding open quantum system. A powerful tool to describe open quantum systems and resonances in the scattering cross section is the scattering matrix theory\cite{bib:Newton, bib:BookMo}. This well established approach\cite{bib:Newton} relates resonances to the poles of the scattering matrix. The resonances similar to the energy levels exhibit crossings and anticrossings\cite{bib:grecchi,bib:Gor} under a variation of the system parameters. However, an exact location of the resonance on the real energy axis coincides with the real part of the S-matrix complex pole only for narrow (i.e. located near the real axis in the complex energy plane) and weakly interacting resonances\cite{bib:Spole1,bib:Newton,bib:Gor}. The Breit-Wigner formula describes exactly this situation. In the case of a wide but still isolated resonance a better fit for the resonance location could be obtained by taking the modulus of the pole complex energy value\cite{bib:AbsEn}. A direct equivalence between the scattering matrix poles and eigenvalues of a Hermitian Hamiltonian with the outgoing Siegert boundary conditions has been established in Ref.\cite{bib:Tolstikhin}. Hence one should expect that the S-matrix language would be appropriate to describe the $\mathcal{PT}$-SB phenomena in $\mathcal{PT}$-symmetric systems as well. The evolution of the S-matrix poles in $\mathcal{PT}$-symmetric systems do bear some features peculiar to the EP picture\cite{bib:rotter1995, bib:Res} but the cross section versus energy dependence doesn't match exactly the evolution of the S-matrix poles\cite{bib:rotter1995, bib:Gor, bib:rotter2015}. It was established in Ref.\cite{bib:Cho} for optical gain/loss media that the eigenvalues of the S-matrix rather than the poles should be used to describe physically observable $\mathcal{PT}$-SB. A pair of unimodular eigenvalues of the S-matrix turns into a pair of nonunimodular values at the scattering matrix EP\cite{bib:Cho}. In a $\mathcal{PT}$-symmetric phase in both electromagnetic modes corresponding to these eigenvalues gain and loss are balanced while in a $\mathcal{PT}$-nonsymmetric phase one mode is attenuated and the other one is amplified. Later it was shown by means of numerical modeling\cite{bib:SmatrEig,bib:Ambi} that in the parameter space there is a line separating $\mathcal{PT}$-symmetric and  $\mathcal{PT}$-nonsymmetric phases. This line can match proper boundary conditions by EP points of the $\mathcal{PT}$-symmetric Hamiltonian. 

A coalescence of three resonances with the unit transparency into a one resonance with the unit transparency has been described in Ref.\cite{bib:rotter1995} in a model of interacting quantum dots. However, in looking for a resonance behavior similar to EP picture one should expect a coalescence of two resonances with the unit transparency and their transformation into a single resonance with a transparency smaller than one. Just this transformation was described in the dissipationless resonant tunneling heterostructure (RTS) in Ref.\cite{bib:Gor}, which was called collapse of resonances (CR). CR is accompanied by SB of the electronic density distribution, which is symmetric at the unit transparency resonance and becomes asymmetric after CR. Later in Ref.\cite{bib:PTScat} CR in a multiwell heterostructure was associated with $\mathcal{PT}$-symmetry breaking in the underlying Schr\"oedinger equation with the $\mathcal{PT}$-symmetric Robin boundary conditions. 

Recently in Ref.\cite{bib:GorSh} it was shown that the exact location of the unity resonances of a spatially symmetric RTS on the real energy axis coincide with the eigenvalues of some auxiliary $\mathcal{PT}$-symmetric non-Hermitian Hamiltonian. CR corresponds to EP of this auxiliary Hamiltonian. $\mathcal{T}$-noninvariance of the auxiliary Hamiltonian describing a dissipationless and hence primarily $\mathcal{T}$-invariant RTS is related to the formulation of the scattering problem with a given direction of the particle flow. The reversal of the particle flow results in sign reversal of terms in auxiliary Hamiltonian responsible for $\mathcal{T}$-symmetry violation. The reversal of the particle flow results also in the spatial inversion of the asymmetric electron distribution at the resonance energy in the $\mathcal{PT}$-symmetry broken phase.

In the present paper we study in details the phenomenon of $\mathcal{PT}$-SB in a dissipationless RTS and establish a direct correspondence between the problem of finding the scattering transparency peaks and some Hamiltonian eigenvalue problem. The structure of the paper is as follows. In Sec.~\ref{Sec.Mod} we introduce a tight-binding model of RTS that makes it possible to obtain basic results in an analytical form. We apply a general approach to quantum transport based on the nonequilibrium Green function (NEGF) formalism and obtain an expression for the RTS transparency, which directly maintains a connection between the poles of the scattering matrix (transparency coefficient) and the eigenvalues of an effective non-Hermitiam Hamiltonian. In Sec.~\ref{Sec.Aux} we show that an auxiliary non-Hermitian Hamiltonian can be introduced, whose eigenvalues  determine $\it{exactly}$ the transparency peak positions. The description of an open quantum system on the basis of the auxiliary non-Hermitiam Hamiltonian is equivalent to the description of the corresponding closed quantum system with the complex valued third type boundary conditions. The equivalence between the resonance peaks location and the eigenvalue problem for the auxiliary Hamiltonian helps to construct a family of non-Hermitiam Hamiltonians, which are not $\mathcal{PT}$-symmetric but possess real eigenvalues. In Sec.~\ref{Sec.Symm} we study a spatially symmetric RTS. The auxiliary non-Hermitian Hamiltonian in this case becomes $\mathcal{PT}$-symmetric and can possess all real eigenvalues, which determine $\it{exactly}$ the positions of the unit transparency peaks. We show that EP of the auxiliary non-Hermitian Hamiltonian in RTS with an even number of wells corresponds to a collapse of resonances accompanied by SB of the electron probability distribution while in RTS with an odd number of wells the spatial symmetry could retain. In Sec.~\ref{Sec.Num} a complimentary description of SB in a spatially symmetric RTS is presented on the basis of a numerical solution of the envelope wavefunction Schr\"odinger equation. In Sec.~\ref{Sec.Cat} we show that a variety of CR in RTS can be described properly in terms of the catastrophe theory\cite{bib:Gilm}. In these terms different  types of CR correspond to different types of catastrophes. The last section contains a summary and a short discussion of the results. 

\section{Model}
\label{Sec.Mod}
The system under study is an arbitrary multi-well resonant-tunneling heterostructure with only a single energy level in each well. Physically this means that we consider only one subband generated by these quasilocalized states and ignore any intersubband interaction. The model describes on equal grounds a multiple quantum dot linear chain coupled to the continuum (similar to the model considered in Ref.\cite{bib:rotter1995, bib:IRot2004}).  The Hamiltonian of this system in the tight-binding approximation, accounting for only the nearest neighbor hopping is expressed as\cite{bib:Car1}:
\begin{equation}
\hat H=\hat H_{0}+\Hat H_{LC}+\Hat H_{RC}.
\label{eqHamilt}
\end{equation}
The first term in~(\ref{eqHamilt}) describes isolated $N$-well RTS:
\begin{equation}
\hat H_{0}=\sum_{i=1}^{N}{\varepsilon_{i}a_{i}^{\dag}a_{i}}+\sum_{i=1}^{N-1}{\left(\tau_{i}a_{i+1}^{\dag}a_{i}+h.c.\right)},
\label{eqH0}
\end{equation}
where $a_{i}^{\dag}(a_{i})$ are the creation (annihilation) operators of the electrons in the $i$-th well with the energy $\varepsilon_{i}$ and $\tau_{i}$ is the tunneling matrix element between the $i$-th and the $(i+1)$-th wells. The Hamiltonians $\Hat H_{LC}/\Hat H_{RC}$ describe the left/right contacts with the spectrum $\varepsilon_{p}$ and the interaction between the system and contacts via the tunneling elements $t_{L(R)}$:
\begin{multline}
\Hat H_{LC(RC)}=\sum_{p}{\varepsilon_{p}a_{L(R),p}^{\dag}a_{L(R),p}}\\
+\sum_{p}{\left(t_{L(R)}a_{1(N)}^{\dag}a_{L(R),p}+h.c.\right)},
\end{multline}
where the operator $a_{L(R),p}$ corresponds to the state with momentum $p$ in the left(right) contact.

We use the Keldysh formalism in the above tight-binding approximation\cite{bib:Car1,bib:Kapa}. We begin with the retarded Green's function of the isolated $N$-well RTS:
\begin{equation}
\hat G^{0r}(\omega)=\left(\omega\hat I-\hat H_{0}\right)^{-1},
\label{eqIsGFDef}
\end{equation}
where $\hat I$ is the identity $N\times N$ matrix. Accounting for the interaction with the contacts we can follow, for example, Ref.\cite{bib:Car1} and write down the full propagator from the $1$-st to the $N$-th well in the form:
\begin{equation}
{G}_{1N}^r=\frac{G_{1N}^{0r}}{\Delta},
\label{eq3}
\end{equation}
with
\begin{equation}
\Delta=(1-\Sigma_{L}G_{11}^{0r})(1-\Sigma_{R}G_{NN}^{0r})-\Sigma_{L}\Sigma_{R}G_{1N}^{0r}G_{N1}^{0r}
\label{delta}
\end{equation}
Here $G_{ij}^{0r}$ are appropriate elements of $\hat G^{0r}$ matrix and $\Sigma_{L,R}$ are the contact self-energies defined as:
\begin{equation}
\Sigma_{L,R}=|t_{L,R}|^{2}g_{L,R}^{r}=\delta_{L,R}-i\Gamma_{L,R},
\label{eqSigma}
\end{equation}
where $g_{L,R}^{r}$ are the retarded Green's functions in the contacts. From Eq.~(\ref{eqIsGFDef}) one can derive the expression for the components of $\hat G^{0r}$ in the form:
\begin{equation}
\begin{split}
G_{11}^{0r}=\frac{D_{0}^{1}}{D_{0}^{0}},\\
G_{1N}^{0r}=\frac{\tau_{1}\cdot...\cdot\tau_{N-1}}{D_{0}^{0}},\\
G_{N1}^{0r}=\frac{\tau_{1}^{*}\cdot...\cdot\tau_{N-1}^{*}}{D_{0}^{0}},\\
G_{NN}^{0r}=\frac{D_{1}^{0}}{D_{0}^{0}}.
\label{eqq8}
\end{split}
\end{equation}
Here we have introduced $D_{q}^{p}$ -- the determinant of the matrix $(\omega\hat I-\hat H_{0})$ with the first $p$ rows and $p$ columns and the last $q$ rows and $q$ columns crossed out. Substituting~(\ref{eqq8}) into~(\ref{eq3}) one gets:
\begin{equation}
{G}_{1N}^r=\frac{\tau_{1}\cdot...\cdot\tau_{N-1}}{D_{0}^{0}-\Sigma_{L}D_{0}^{1}-\Sigma_{R}D_{1}^{0}+\Sigma_{L}\Sigma_{R}K},
\label{A6}
\end{equation}
where
\begin{equation*}
K=\frac{1}{D_{0}^{0}}(D_{0}^{1}D_{1}^{0}-|\tau_{1}|^{2}\cdot...\cdot|\tau_{N-1}|^{2}).
\end{equation*}
The matrix $\omega\hat I-\hat H_{0}$ is tridiagonal, so we can use the following recurrence relation:
\begin{equation}
D_{q}^{p}=(\omega-\varepsilon_{p+1})D_{q}^{p+1}-|\tau_{p+1}|^{2}D_{q}^{p+2}.
\label{A7}
\end{equation}
Let us consider the expression $D_{0}^{0}D_{1}^{1}-D_{0}^{1}D_{1}^{0}$. With the help of~(\ref{A7}) we get:
\begin{widetext}
\begin{multline}
D_{0}^{0}D_{1}^{1}-D_{0}^{1}D_{1}^{0}=|\tau_{1}|^{2}\left(D_{0}^{1}D_{1}^{2}-D_{0}^{2}D_{1}^{1}\right)=...=|\tau_{1}|^{2}\cdot...\cdot|\tau_{N-2}|^{2}\left(
\begin{vmatrix}
\omega-\varepsilon_{N-1} & -\tau_{N-1}\\
-\tau_{N-1}^{*} & \omega-\varepsilon_{N}
\end{vmatrix}
-(\omega-\varepsilon_{N-1})(\omega-\varepsilon_{N})\right)\\
=-|\tau_{1}|^{2}\cdot...\cdot|\tau_{N-1}|^{2}.
\label{A8}
\end{multline}
\end{widetext}
From~(\ref{A8}) it is straightforward that $K$ in~(\ref{A6}) is simplified to $K=D_{1}^{1}$. Thus, the denominator in~(\ref{A6}) is nothing but the determinant of $\omega\hat I-\hat H_{eff}$:
\begin{equation}
{G}_{1N}^r=\frac{\tau_{1}\cdot...\cdot\tau_{N-1}}{\det{(\omega\hat I-\hat H_{eff})}},
\label{A9}
\end{equation}
where $\hat H_{eff}$ is the effective Hamiltonian of the open system under consideration (RTS interacting with the continuum in the contacts):
\begin{equation}
\hat H_{eff}=\hat H_{0}+\hat H_{L}+\hat H_{R}.
\label{eqHeff}
\end{equation}
Here the interaction with the contacts is taken into account by $(\hat H_{L})_{ij}=\Sigma_{L}\delta_{i1}\delta_{j1}$ and $(\hat H_{R})_{ij}=\Sigma_{R}\delta_{iN}\delta_{jN}$. Real parts $\delta_{L,R}$ of the self-energies~(\ref{eqSigma}) correspond to the energy shift and imaginary parts $\Gamma_{L,R}$ describing the decay into the bulk's continuum, which is close in a sense to the Feshbach optical potential\cite{bib:F1,*bib:F2}. The matrix $H_{eff}$ takes the form:
\begin{equation}
\hat H_{eff}=\begin{pmatrix}
		\varepsilon_{1}+\delta_{L}-i\Gamma_{L} & \tau_{1} & \ldots & 0 & 0\\
		\tau_{1}^{*} & \varepsilon_{2} & \ldots & 0 & 0\\
		\vdots & \vdots & \ddots & \vdots & \vdots\\
		0 & 0 & \ldots & \varepsilon_{N-1} & \tau_{N-1}\\
		0 & 0 & \ldots & \tau_{N-1}^{*} &\varepsilon_{N}+\delta_{R}-i\Gamma_{R}
		\end{pmatrix}.
\label{eqHeffM}
\end{equation}

In the Keldysh formalism the current through the structure consisting of $N$ coupled wells can be written in the standard form\cite{bib:Car1}:
\begin{equation}
I=\frac{e}{2\pi}\int{T_{NW}(\omega)(f_{L}(\omega)-f_{R}(\omega))d\omega}.
\label{eq0}
\end{equation}
where $f_{L,R}$ is the Fermi distribution function in the left/right contact. The transmission probability is directly expressed through the full propagator ${G}_{1N}^r$:
\begin{equation}
T_{NW}=4\Gamma_{L}\Gamma_{R}|{G}^{r}_{1N}|^{2}.
\label{eq01}
\end{equation}
Hence if we substitute~(\ref{A9}) into~(\ref{eq01}), we get for the transmission coefficient the well known expression -- a fraction with the characteristic polynomial of the system's effective Hamiltonian in the denominator\cite{bib:Cel,bib:Celardo2010}:
\begin{equation}
T_{NW}=\frac{P^{2}}{\left|\det\left(\omega\hat I-\hat H_{eff}\right)\right|^{2}},
\label{eqTNclass}
\end{equation}
where $P^{2}=4\Gamma_{L}\Gamma_{R}|\tau_{1}|^2\cdot...\cdot|\tau_{N-1}|^{2}$.
From~(\ref{eqTNclass}) it follows immediately that the poles of the transmission probability and hence the poles of the scattering matrix coincide with the eigenvalues of the non-Hermitian Hamiltonian  $\hat H_{eff}$~(\ref{eqHeff}) and (\ref{eqHeffM}), which are complex numbers. For $N=1$ Eqs.~(\ref{eqHeffM}) and~(\ref{eqTNclass}) give the familiar Breit-Wigner formula. The resonance location on the real energy axis in this case exactly matches the real part of the transmission probability (as well as the S-matrix) pole. However, for strongly interacting resonances, which could undergo a coalescence, another approach is required. 

\section{Auxiliary Hamiltonian}
\label{Sec.Aux}
Let us consider the denominator of~(\ref{eqTNclass}), where we substitute $\Sigma_{L,R}$ as $\delta_{L,R}-i\Gamma_{L,R}$ correspondingly. One can expand the determinant and rewrite it in the following way: 
\begin{equation}
\det{(\omega\hat I-\hat H_{eff})}=\tilde D_{0}^{0}+i\Gamma_{L}\tilde D_{0}^{1}+i\Gamma_{R}\tilde D_{1}^{0}-\Gamma_{L}\Gamma_{R}\tilde D_{1}^{1},
\label{B1}
\end{equation}
where $\tilde D_{q}^{p}$ differs from $D_{q}^{p}$ by taking into account the shifts $\delta_{L,R}$ in the energy of the outer wells, nevertheless, Eq.~(\ref{A8}) is also correct for $\tilde D_{q}^{p}$. Quantities $\tilde D_{q}^{p}$ are real for real values of $\omega$ and real energies $\varepsilon_{i}$, because the only possible complex matrix elements $\tau_{i}$ appear in the expression for $\tilde D_{q}^{p}$ multiplied by their complex conjugates. Thus, if we add and subtract $P^{2}=4\Gamma_{L}\Gamma_{R}|\tau_{1}|^{2}\cdot...\cdot|\tau_{N-1}|^{2}$ to the squared modulus of~(\ref{B1}), we can rewrite it as:
\begin{widetext}
\begin{multline}
\left|\det{(\omega\hat I-\hat H_{eff})}\right|^{2}=\left|\tilde D_{0}^{0}+i\Gamma_{L}\tilde D_{0}^{1}+i\Gamma_{R}\tilde D_{1}^{0}-\Gamma_{L}\Gamma_{R}\tilde D_{1}^{1}\right|^{2}-P^{2}+P^{2}=\left(\tilde D_{0}^{0}-\Gamma_{L}\Gamma_{R}\tilde D_{1}^{1}\right)^{2}+\left(\Gamma_{L}\tilde D_{0}^{1}+\Gamma_{R}\tilde D_{1}^{0}\right)^{2}\\
+4\Gamma_{L}\Gamma_{R}\left(\tilde D_{0}^{0}\tilde D_{1}^{1}-\tilde D_{0}^{1}\tilde D_{1}^{0}\right)+P^{2}=|Q|^{2}+P^{2},
\label{B2}
\end{multline}
\end{widetext}
where 
\begin{equation}
|Q|^{2}=(\tilde D_{0}^{0}+\Gamma_{L}\Gamma_{R}\tilde D_{1}^{1})^{2}+(\Gamma_{L}\tilde D_{0}^{1}-\Gamma_{R}\tilde D_{1}^{0})^{2},
\label{Q1}
\end{equation}
This equation determines $Q$ up to an arbitrary phase factor. We can use this degree of freedom to define $Q$ as
\begin{equation}
Q=\tilde D_{0}^{0}-i\Gamma_{L}\tilde D_{0}^{1}+i\Gamma_{R}\tilde D_{1}^{0}+\Gamma_{L}\Gamma_{R}\tilde D_{1}^{1},
\label{Q2}
\end{equation}
Comparing Eqs.~(\ref{B1}) and~(\ref{Q2}) it is easily seen that $Q$ in~(\ref{Q1}) and~(\ref{Q2}) is the characteristic polynomial of some auxiliary Hamiltonian $\hat{H}_{aux}$:
\begin{equation}
Q=\det{\left(\omega\hat I-\hat{H}_{aux}\right)}
\label{eqQd}
\end{equation}
with
\begin{equation}
\hat{H}_{aux}=\hat H_{0}+\hat H_{L}^{*}+\hat H_{R}.
\label{eqHeff2}
\end{equation}
Here $\hat H_{L}^{*}$ is the complex conjugate of $\hat H_{L}$ from~(\ref{eqHeff}). Definition~(\ref{eqHeff2}) describes the electron flow from the left to the right. On the other hand, the opposite electron flow direction leads to the definition $\hat{H}_{aux}=\hat H_{0}+\hat H_{L}+\hat H_{R}^{*}$. As one can expect, the transmission coefficient is insensitive to the electron flow direction. The matrix of the auxiliary Hamiltonian is
\begin{equation}
\hat H_{aux}=\begin{pmatrix}
		\varepsilon_{1}+\delta_{L}+i\Gamma_{L} & \tau_{1} & \ldots & 0 & 0\\
		\tau_{1}^{*} & \varepsilon_{2} & \ldots & 0 & 0\\
		\vdots & \vdots & \ddots & \vdots & \vdots\\
		0 & 0 & \ldots & \varepsilon_{N-1} & \tau_{N-1}\\
		0 & 0 & \ldots & \tau_{N-1}^{*} &\varepsilon_{N}+\delta_{R}-i\Gamma_{R}
		\end{pmatrix}.
\label{eqHeff2M}
\end{equation}

Ergo we have shown that the transmission of the arbitrary multi-well system~(\ref{eqTNclass}) can be rewritten as:
\begin{equation}
T_{NW}=\frac{P^{2}}{|Q|^{2}+P^{2}},
\label{eqTN}
\end{equation}
with $Q=\det{(\omega\hat I-\hat{H}_{aux})}$ that is the characteristic polynomial of the non-Hermitian auxiliary Hamiltonian describing the electron flow from the left to the right.  

\begin{figure}
\includegraphics{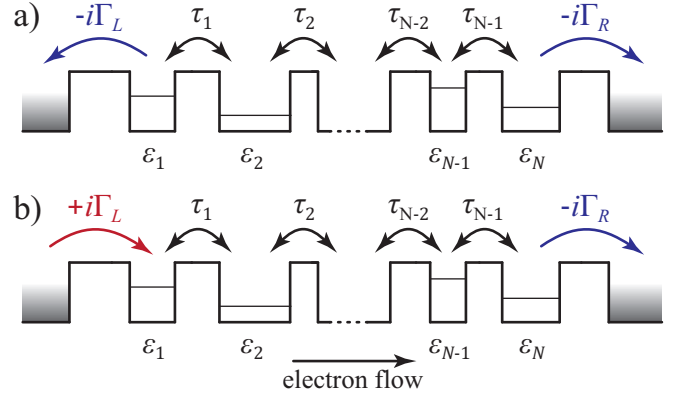}
\caption{\label{fig1}(Color online) Illustrative interpretation of the effective Hamiltonian (a) and the auxiliary Hamiltonian (b) of an arbitrary $N$-well resonant-tunneling heterostructure sandwiched between two bulk contact regions. The horizontal thin solid line in each well represents the bare quantum well energy level $\varepsilon_{i}$.}
\end{figure}

Thus, we get, that the resonances of the transmission are {\it exactly} determined by the eigenvalues of the auxiliary Hamiltonian, which is the main result of this section. Formally the auxiliary Hamiltonian $\hat H_{aux}$ differs from $\hat H_{eff}$ in that it has the opposite signs of the imaginary terms in the first and the last element on the main diagonal (Fig.~\ref{fig1}). These terms in $\hat H_{aux}$ correspond to inflow and outflow of electrons. The imaginary terms in the Hamiltonian allowing for a description of incoming and outgoing electrons were considered also in Refs.\cite{bib:Res, bib:Jin2010}. In $\hat H_{eff}$ the imaginary terms making the Hamiltonian non-Hermitian describe only the outflow of electrons that characterize a transformation of a closed quantum system to an open one, whose eigenstates acquire a finite lifetime. 
 
Expression~(\ref{eqTN}) represents a compact generalization of the Breit-Wigner formula to the multilevel case. The real roots of $Q$ are {\it exactly} the positions of the unity peaks of the transmission. The complex roots correspond to resonances with a transmission value smaller than unity. In the case of well separated non-unity resonances the positions of non-unity peaks are approximately determined by the reals parts of the complex roots of $Q$.

Interrelation between the auxiliary non-Hermitian Hamiltonian and the underlying Hermitian Hamiltonian of the closed system can be easily understood in the following way. Let's consider dynamics of tight-binding site (quantum well or quantum dot) amplitudes $a_{i}$ governed by $\hat H_{aux}$: 
\begin{eqnarray}
\label{eqGenSite}
\omega a_{i}&=&\varepsilon_{i}a_{i}-\tau_{i-1}^{*}a_{i-1}-\tau_{i}a_{i} \quad 1<i<N,\\
\label{eqFS}
\omega a_{1}&=&(\varepsilon_{1}+\delta_{L}+i\Gamma_{L})a_{1}-\tau_{1}a_{2},\\
\label{eqLS}
\omega a_{N}&=&(\varepsilon_{1}+\delta_{R}-i\Gamma_{R})a_{N}-\tau_{N-1}^{*}a_{N-1}.
\end{eqnarray}
Eqns.~(\ref{eqFS}-\ref{eqLS}) can be considered as complex boundary conditions (BC) in the site representation for the Hermitian Hamiltonian describing the closed quantum system. So, the system described by the auxiliary Hamiltonian is a bounded system with special BC. To derive BC in a standard form we add two outer sites $i=0,N+1$ with extrapolated from inner region amplitudes ($\tilde a_{0}$ and $\tilde a_{N+1}$). The extrapolated amplitudes satisfy: 
\begin{equation}
\begin{split}
\tau_{0}^{*}\tilde a_{0}&=-i\Gamma_{L}a_{1},\\
\tau_{N}\tilde a_{N+1}&=i\Gamma_{R}a_{N},
\end{split}
\label{eqBC}
\end{equation}
which makes equation~(\ref{eqGenSite}) applicable in the whole inner region $1\leq i\leq N$.

The standard form of BC can be obtained  treating the site amplitudes $a_{i}$ as slowly varying functions of the site number, which we can express in terms of the envelope function $\psi(x)$ of the continuous coordinate variable $x$:

\begin{equation}
\begin{split}
\tilde a_{0}\approx A\psi(0),&\quad a_{1}\approx A(\psi(0)+d_{L}\partial_{x}\psi(0)),\\
\tilde a_{N+1}\approx A\psi(L),&\quad a_{N}\approx A(\psi(L)-d_{R}\partial_{x}\psi(L)).
\end{split}
\label{eqEnvel}
\end{equation}
Here $A$ is a normalization constant (a conversion factor from the site representation into the continuous one) and $d_{L,R}$ -- intersite spacing.  Substituting (\ref{eqEnvel}) into Eq.~(\ref{eqBC}), we get mixed (third type) BC for the envelope functions:
\begin{equation}
-\partial_{x}\psi(0)=\lambda_{L}\psi(0),\quad\partial_{x}\psi(L)=\lambda_{R}\psi(L),
\label{eqBCEn}
\end{equation}
where
\begin{equation}
\lambda_{L}=-i\frac{\tau_{0}^{*}+i\Gamma_{L}}{\Gamma_{L}d_{L}},\quad\lambda_{R}=i\frac{\tau_{N}-i\Gamma_{R}}{\Gamma_{R}d_{R}}.
\label{eqLam}
\end{equation}
Thus, the auxiliary Hamiltonian allows to map the unbounded scattering problem for RTS of a general form (including nonsymmetric RTS) into a bounded one with certain BC~(\ref{eqBCEn}). This result generalizes the observation of Ref.~\cite{bib:PTScat} that the collapse of the overbarrier resonances in a symmetric multiple quantum well structure can be described on the basis of a Hermitian Schr\"oedinger equation with the $\mathcal{PT}$-symmetric Robin boundary conditions.A similar feature was described in Ref.\cite{bib:Ambi} in studying the S-matrix eigenvalues properties of $\mathcal{PT}$-symmetric optical systems. However, the parameter $\lambda_{L,R}$ in our case is a complex quantity that is different in general on different subsets of the boundary, while in Ref.\cite{bib:Ambi} it is pure imaginary and takes the same value on the whole boundary (Robin BC). It reflects the fact that we are dealing with a scattering problem at a given direction of the particle flow but not with the scattering matrix eigenvalue problem.

To demonstrate efficiency of the auxiliary Hamiltonian notion consider perfect transmission of non-symmetric RTS.  Any RTS can be considered as a two-barrier RTS with composite barriers. Perfect transmission takes place when the transparencies of the composite barriers are equal to each other. According to the results of this section the unit transparency is achieved at the energies, which are the real eigenvalues of the non-Hermitian Hamiltonian. However, a non-symmetric RTS is characterized by an auxiliary non-Hermitian Hamiltonian which is, generally, $\it {not}$ $\mathcal{PT}$-symmetric. Hence we obtain the method of constructing non-Hermitian Hamiltonians with real eigenvalues, which do not possess $\mathcal{PT}$-symmetry. 

Consider as an example an asymmetric double-well structure. In the case of a perfect trasmission the polinomial $Q$ of the structure should have one real and one complex root:
\begin{equation}
Q=(\omega-a)(\omega-b-ic),
\label{eqNeedQ}
\end{equation}
where $a$, $b$ and $c$ are real. On the other hand, {\bf a} general form of $Q_{gen}$ for an asymmetric structure is:
\begin{equation}
Q_{gen}=\begin{vmatrix}
	\omega-\varepsilon_{1}-i\Gamma_{L} & -\tau_{1}\\
	-\tau^{*}_{1} & \omega-\varepsilon_{2}+i\Gamma_{R}
\end{vmatrix}.
\label{eqGenQ}
\end{equation}
Expanding~(\ref{eqGenQ}) and comparing it's coefficients with~(\ref{eqNeedQ}) one can conclude that it is possible to make one unity transmission peak by tuning only one parameter, for example, setting $\varepsilon_{1}$ to be:
\begin{equation}
\varepsilon_{1}=\varepsilon_{2}\pm(\Gamma_{L}-\Gamma_{R})\sqrt{\frac{|\tau_{1}|^{2}}{\Gamma_{L}\Gamma_{R}}-1}.
\label{eqE1}
\end{equation}
This substitution makes transmission to be perfect at
\begin{equation}
\omega_{perf}=\varepsilon_{2}\mp\sqrt{\frac{|\tau_{1}|^{2}\Gamma_{R}}{\Gamma_{L}}-\Gamma_{R}^{2}}.
\label{eqWPerf}
\end{equation}
The sign choice in~(\ref{eqWPerf}) depends on the sign chosen in~(\ref{eqE1}). Thus, we have shown that a double-well asymmetric structure can have perfect transmission at any given energy $\omega_{0}$. This can be achieved by tuning system's parameters in a way to have $\omega_{0}=\omega_{perf}$ from~(\ref{eqWPerf}) and setting $\varepsilon_{1}$ in accordance with~(\ref{eqE1}). In turn the non-$\mathcal{PT}$-symmetric non-Hermitian Hamiltonian described by matrix~(\ref{eqGenQ}) possesses real eigenvalues Q.E.D.

\section{symmetry breaking in symmetric structures}
\label{Sec.Symm}
In symmetric RTS ($\Gamma_{L}=\Gamma_{R}=\Gamma$, $\varepsilon_{N+1-i}=\varepsilon_{i}$, $\delta_{L}=\delta_{R}$ and $\tau_{N-i}=\tau_{i}$) the auxiliary Hamiltonian~(\ref{eqHeff2}) becomes $\mathcal{PT}$-symmetric:
\begin{equation}
\hat{H}_{aux}^{symm}=\hat H_{\mathcal{PT}}=\hat H_{0}+\hat H_{\Gamma},
\label{eqHPT}
\end{equation}
where $(\hat H_{\Gamma})_{ij}=i\Gamma(\delta_{i1}\delta_{j1}-\delta_{iN}\delta_{jN})$ corresponds to the case of the electron flow from the left to the right. Thus, for symmetric RTS, the unity peaks of transmission, according to~(\ref{eqTN}), are defined by the eigenvalues of the non-Hermitian $\mathcal{PT}$-symmetric Hamiltonian~(\ref{eqHPT}), which can be real and possess EP. Under these assumptions the mixed BC~(\ref{eqBCEn}) become Robin $\mathcal{PT}$-symmetric with:
\begin{equation*}
\lambda_{L}= \lambda_{R}^{*}=\lambda.
\end{equation*}
Hereinafter we assume the states in each well to have the same energy ($\varepsilon_{i}=\varepsilon_{0}$ for each $i$) and the energy shifts in the outer wells to be zero ($\delta_{L}=\delta_{R}=0$). These restrictions make it possible to describe the collapse formally of an arbitrary number of resonances. The assumption about the real parts of the self-energies ($\delta_{L}=\delta_{R}=0$) physically means that we consider energy levels of RTS located near the center of the band in contacts\cite{bib:RealSE2} (equivalently we can treat the contact band to be wide compared to the RTS subband\cite{bib:RealSE1}). 

In the case of symmetric structures all the coefficients of polynomial $Q$ are real and it has some remarkable properties, which are rigorously proved in Appendix~\ref{ApA}. It can be shown that $Q$ is the $N/2$-th order polynomial of $(\omega-\varepsilon_{0})^{2}$ for even $N$ and $(N-1)/2$-th order polynomial of $(\omega-\varepsilon_{0})^{2}$ multiplied by $(\omega-\varepsilon_{0})$ for odd $N$. Hence all the roots of $Q$ are symmetric in pairs with respect to the energy $\omega=\varepsilon_{0}$. There are enough independent system's parameters $\Gamma$ and $\{\tau_{i}\}$ to control the positions of all the roots of $Q$ (taking into account the aforementioned restrictions). The coalescence of two real roots of $Q$ at EP and then their transformation into a pair of complex conjugates corresponds to the coalescence of two unity resonances into one with a transmission value smaller than unity. In structures with an even number of wells $N$ all the roots of $Q$ (all the eigenvalues of $\mathcal{PT}$-symmetric $\hat H_{aux}$) can be made complex under a variation of the Hamiltonian parameters, which means the absence of unity transmission peaks. At the resonance with the transparency smaller than unity the spatial symmetry is broken (see this section below). Hence, a coalescence of resonances in such a structure with an even number of wells means $\mathcal{PT}$-symmetry breaking of the auxiliary Hamiltonian. It is the collapse of resonances (CR), which was described in Ref.\cite{bib:Gor} for $N=2$ resonances. However, for an odd $N$ there is always at least one real root $\omega=\varepsilon_{0}$ (see Appendix~\ref{ApA} for details). Hence, there is always at least one perfect transmission resonance, which preserves the spatial symmetry. For $N=3$ resonances it was observed in Ref.\cite{bib:IRot2004}. Generally, the interaction of pairs of eigenvalues is the only type of $\hat H_{aux}$ eigenvalues interaction\cite{bib:IRot1,bib:IRot2}. However, under a specific choice of the Hamiltonian parameters (just as made in this section) all the unity resonances can coalesce together forming an $N$-th order exceptional point (EP) of $\hat H_{aux}^{symm}=\hat H_{\mathcal{PT}}$ (they do it in pairs but at the same values of the parameters and energy).

Consider the electron's density distribution in a symmetric RTS. In the absence of incoherent effects (inelastic scattering, etc.) one can write the electron concentration in the $i$-th well in the Keldysh formalism as\cite{bib:ElDen}:
\begin{equation}
n_{i}=\frac{\Gamma}{2\pi}\left(f_{L}|G_{i1}^{r}|^{2}+f_{R}|G_{iN}^{r}|^{2}\right),
\label{ni}
\end{equation}
where $f_{L,R}$ is the Fermi-Dirac distribution function in the left/right contact correspondingly. The propagators $G_{i1}^{r}$ and $G_{iN}^{r}$ can be obtained from the following Dyson equation\cite{bib:Car1}:
\begin{equation}
G_{ij}^{r}=G_{ij}^{0r}+\Sigma_{L}G_{i1}^{0r}G_{1j}^{r}+\Sigma_{R}G_{iN}^{0r}G_{Nj}^{r},
\label{Dys}
\end{equation}
where the Green's function of the isolated system $G_{ij}^{0r}$ is derived from~(\ref{eqIsGFDef}). The self-energies $\Sigma_{L,R}=\delta_{L,R}-i\Gamma_{L,R}$ are simplified to $\Sigma_{L}=\Sigma_{R}=-i\Gamma$ under the assumptions made in this section. From~(\ref{Dys}) we can get:
\begin{equation}
\begin{split}
G_{i1}^{r}=\frac{1}{\Delta}\left[G_{i1}^{0r}+i\Gamma\left(G_{NN}^{0r}G_{i1}^{0r}-G_{iN}^{0r}G_{N1}^{0r}\right)\right],\\
G_{iN}^{r}=\frac{1}{\Delta}\left[G_{iN}^{0r}+i\Gamma\left(G_{11}^{0r}G_{iN}^{0r}-G_{i1}^{0r}G_{1N}^{0r}\right)\right],
\label{Gre}
\end{split}
\end{equation}
with $\Delta$ determined in~(\ref{delta}).

One can see from Eq.~(\ref{ni}), that the electron concentration in the $N-i+1$-th well (symmetric to $i$-th one) is defined by the absolute values of the propagators $G_{N-i+1,1}^{r}$ and $G_{N-i+1,N}^{r}$, which, according to the symmetry of the structure, are equal to the absolute values of $G_{iN}^{r}$ and $G_{i1}^{r}$ correspondingly. So, the difference of the electron concentrations in the symmetric $i$-th and $N-i+1$-th wells can be written as follows:
\begin{equation}
n_{i}-n_{N-i+1}=\frac{\Gamma}{2\pi}\left(f_{L}-f_{R}\right)\left(|G_{i1}^{r}|^{2}-|G_{iN}^{r}|^{2}\right).
\label{niDif}
\end{equation}
Using~(\ref{eqIsGFDef}) one can calculate all $G_{ij}^{0r}$ required in~(\ref{Gre}) in terms of $D_{q}^{p}$ (which is $D_{q}^{p}=D_{p}^{q}$ in a symmetric structure) and then express the quantity $|G_{i1}^{r}|^{2}-|G_{iN}^{r}|^{2}$ in the factored form:
\begin{widetext}
\begin{equation}
|G_{i1}^{r}|^{2}-|G_{iN}^{r}|^{2}=\frac{(D_{0}^{i})^{2}|\tau_{1}|^{2}\cdot...\cdot|\tau_{i-1}|^{2}-(D_{N-i+1}^{0})^{2}|\tau_{i}|^{2}\cdot...\cdot|\tau_{N-1}|^{2}}{(D_{0}^{0})^{4}|\Delta|^{2}}\left[(D_{0}^{0})^{2}+\Gamma^{2}\left((D_{0}^{1})^{2}-|\tau_{1}|^{2}\cdot...\cdot|\tau_{N-1}|^{2}\right)\right].
\label{GreDif}
\end{equation}
\end{widetext}
The last factor in~(\ref{GreDif}), according to~(\ref{A8}) and definition of the polynomial $Q$, is equal to $D_{0}^{0}Q$. So, we can rewrite~(\ref{niDif}) as:
\begin{widetext}
\begin{equation}
n_{i}-n_{N-i+1}=\frac{\Gamma}{2\pi}\left(f_{L}-f_{R}\right)\frac{(D_{0}^{i})^{2}|\tau_{1}|^{2}\cdot...\cdot|\tau_{i-1}|^{2}-(D_{N-i+1}^{0})^{2}|\tau_{i}|^{2}\cdot...\cdot|\tau_{N-1}|^{2}}{(D_{0}^{0})^{3}|\Delta|^{2}}\times Q.
\label{niDif2}
\end{equation}
\end{widetext}
The key feature of Eq.~(\ref{niDif2}) is that ${n_{i}-n_{N-i+1}\propto Q}$. A real energy $\omega_{r}$, corresponding to a unity-valued resonance is equal to a real root of $Q$, so in such a resonance the value of $Q(\omega_{r})$ and, consequently, the value of the difference $n_{i}(\omega_{r})-n_{N-i+1}(\omega_{r})$ is zero for any $i$ -- i.e. the electrons are distributed symmetrically. However, for the case of a resonance with a transmission coefficient value smaller than $1$ (for example, in the postcollapse\cite{bib:GorSh} state), the value of $Q$ at the resonance's position energy $\omega_{c}$ is nonzero and so the difference $n_{i}(\omega_{c})-n_{N-i+1}(\omega_{c})\neq0$, which means the symmetry breaking of the electron density distribution. Thus, in structures with an even number of wells, there is a symmetry breaking accompanying CR, because the transmission at the remaining peak after CR is less than unity. In the case of an odd number of wells, as we have shown above, the remaining peak is always unity-valued and, so, no symmetry breaking happens.

Introduction of the auxiliary Hamiltonian $\hat{H}_{aux}$ makes it possible to perform a direct comparison between the poles of the scattering matrix and location of the transmission resonances by studying the eigenvalues of $\hat{H}_{eff}$ and $\hat{H}_{aux}$. In the tight-binding approximation the poles of S-matrix of a double-well symmetric structure are:
\begin{equation}
\omega_{Spole}^{2W}=\varepsilon_{0}\pm|\tau_{1}|-i\Gamma.
\label{eqSpole}
\end{equation}
The transmission peaks are located at energies, which are eigenvalues of $\hat{H}_{aux}$:
\begin{equation}
\omega_{peak}^{2W}=\varepsilon_{0}\pm\sqrt{|\tau_{1}|^{2}-\Gamma^{2}}.
\label{eqPeak}
\end{equation}
It is easily seen from (\ref{eqSpole}-\ref{eqPeak}), that in the limit of a narrow well separated resonances ($\Gamma\ll\tau_{1}$) positions of the transmission peaks $\omega_{peak}^{2W}$ coincide with the real parts of the S-matrix poles up to the terms linear in $\Gamma/|\tau_{1}|$:
\begin{multline*}
\omega_{peak}^{2W}=\varepsilon_{0}\pm|\tau_{1}|+O\left(\left(\frac{\Gamma}{|\tau_{1}|}\right)^{2}\right)\\
=\R{\omega_{Spole}^{2W}}+O\left(\left(\frac{\Gamma}{|\tau_{1}|}\right)^{2}\right).
\end{multline*}
In the case of an overlapping and wide resonances, when $\Gamma/\tau_{1}$ is not small, the approximation of peaks positions with real parts of S-matrix poles is completely unsuitable. Moreover, the scattering matrix poles do not describe the behavior of overlapping resonances even qualitatively. Indeed, the transmission peaks positions~(\ref{eqPeak}) can coalesce at $\Gamma=|\tau_{1}|$, whereas the real parts of the S-matrix poles~(\ref{eqSpole}) do not coalesce at all. Although, the real parts of the S-matrix poles can exhibit a coalescence as well. For example, in an $N=3$ well structure the S-matrix poles are: $\omega_{Spole}^{3W}=\varepsilon_{0}-i\Gamma+n\sqrt{8|\tau_{1}|^{2}-\Gamma^{2}}$, where $n=\{-1,0,1\}$. The transmission peaks in this case are located at $\omega_{peak}^{3W}=\varepsilon_{0}+n\sqrt{2|\tau_{1}|^{2}-\Gamma^{2}}$, where again $n=\{-1,0,1\}$.

\section{Envelope wavefunction approximation}
\label{Sec.Num}
\begin{figure}
\includegraphics{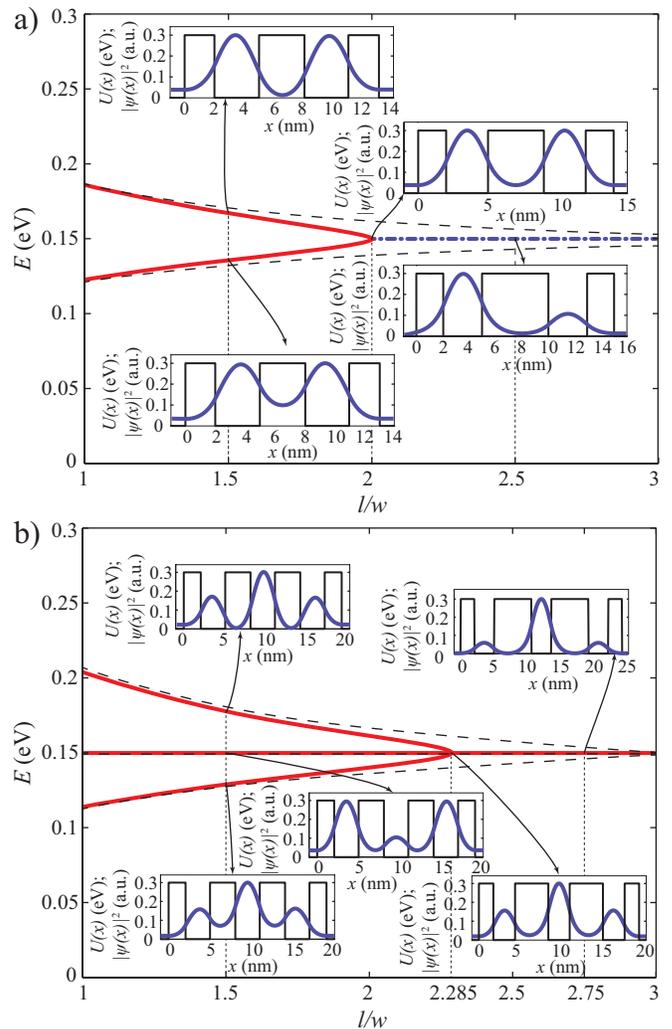}
\caption{\label{fig2}(Color online) Numerically calculated positions of the transmission maxima and real parts of the S-matrix poles vs. the ratio of the central to the outside barrier widths $l/w$ for $N=2$ wells (a) and $N=3$ wells (b) structures. The solid lines correspond to unity valued maxima, the dot-dashed lines -- to non unity maxima peaks and the dashed lines stand for the real parts of the S-matrix poles. Also the electron wavefunction distribution profiles are shown at the resonances at various values of parameter $l/w$.}
\end{figure}

\begin{figure}
\includegraphics{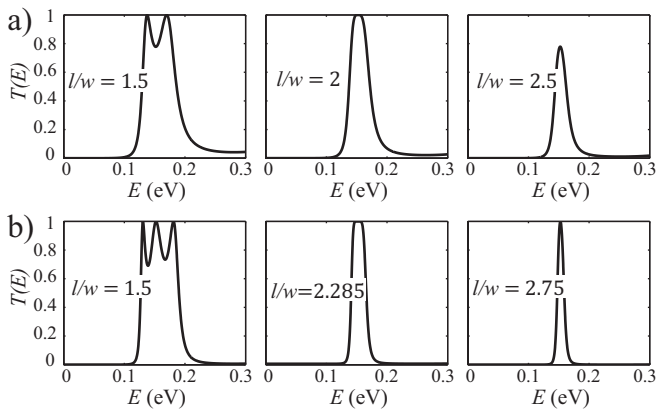}
\caption{\label{fig3}Numerically calculated transmission coefficient vs. the energy plots for $N=2$ wells (a) and $N=3$ wells (b) structures at different values of the central to the outside barrier widths ratio $l/w$, at which the electron's wavefunction profiles in Fig.~{\ref{fig2}} were plotted.}
\end{figure}

The previous discussion was based on a tight-binding approximation with the nearest neighbors hopping. Nevertheless, qualitative conclusions are meaningful beyond this approximation and are quite universal. In this section we demonstrate this universality by means of a numerical solution of an effective mass 1D Shr\"odinger equation: 
\begin{equation}
\left(-\frac{\hbar^2}{2m^{*}}\nabla^2+U(x)\right)\psi(x)=E\psi(x),
\label{eqShr}
\end{equation}
with a step-like potential $U(x)$ describing an $N$-well RTS. To solve Eq.~(\ref{eqShr}) numerically we use the transfer matrix technique\cite{bib:TM,bib:bookTM,bib:TMallDim}. The assumption made in the tight-binding model about the energy shifts: $\delta_{L}=\delta_{R}=0$ {\bf is} taken into account in the Shr\"odinger equation formalism by placing the energy of the quasilocolized states in the middle of the barrier height\cite{bib:Cel}. Figure~\ref{fig2} shows the positions of the transmission maxima vs. the ratio of the central to the outside barrier widths ($l/w$) and the electron wavefunction distribution profiles for the $N=2$ wells and $N=3$ wells structures with the barrier height of $V=0.3$ eV, effective mass $m^{*}=0.067 m_{0}$, fixed outside barrier thickness of $w=2$ nm and wells width of $a=3.0485$ nm (which sets the energy of the quasilocolized states in the middle of the barrier height). The transmission coefficient vs. the energy plots are shown in Fig.~\ref{fig3} for the same values of the $l/w$ ratios as the electron wavefunction profiles in Fig.~\ref{fig2}. In the case of an odd $N$ (Fig.~\ref{fig2}b) the postcollapse transmission is perfect and the symmetry of the electron wavefunction distribution is conserved, whereas for an even $N$ (Fig.~\ref{fig2}a) the postcollapse transparency peak has a non-unity magnitude and the symmetry is broken. 

The physical mechanism of the different behavior of symmetric systems with odd and even numbers of the quantum wells at EP is as follows. As was pointed above and in Refs.\cite{bib:IRot1,bib:IRot2} the coalescence of resonances is a pair process. The  electronic state is degenerate at the EP (contrary to the crossing point) and is formed by a linear combination of the states at the resonances, which undergo the coalescence. In the case of an even-$N$ structure such states are always either adjacent states on the energy axis or they are separated by an even number of resonances (for higher order EP), i.e. they possess the reverse symmetry. Hence, a combination of symmetric and antisymmetric wavefunctions results in the formation of a nonsymmetric wavefunction that is responsible for the spatial symmetry breaking. In the case an odd order EP, which is depicted in Fig.~\ref{fig2}b, a pair of resonances undergoing a coalescence is separated on the energy axis by an odd number of resonances and the corresponding wavefunctions possess the same symmetry. Hence, state at EP preserves its symmetry, which remains unbroken.

A numerical simulation of a triple-well structure shows that both the S-matrix poles and the resonances positions can coalesce, however, at completely different ratios of the parameters in accordance with the analytical tight-binding results of the previous section. Figure~{\ref{fig2}b} with the results of the numerical simulation illustrates this statement -- transmission peaks (solid lines) coalesce at appr.~$l/w=2.285$ and the S-matrix poles (dashed lines) at appr.~$l/w=3$.

\begin{figure}
\includegraphics{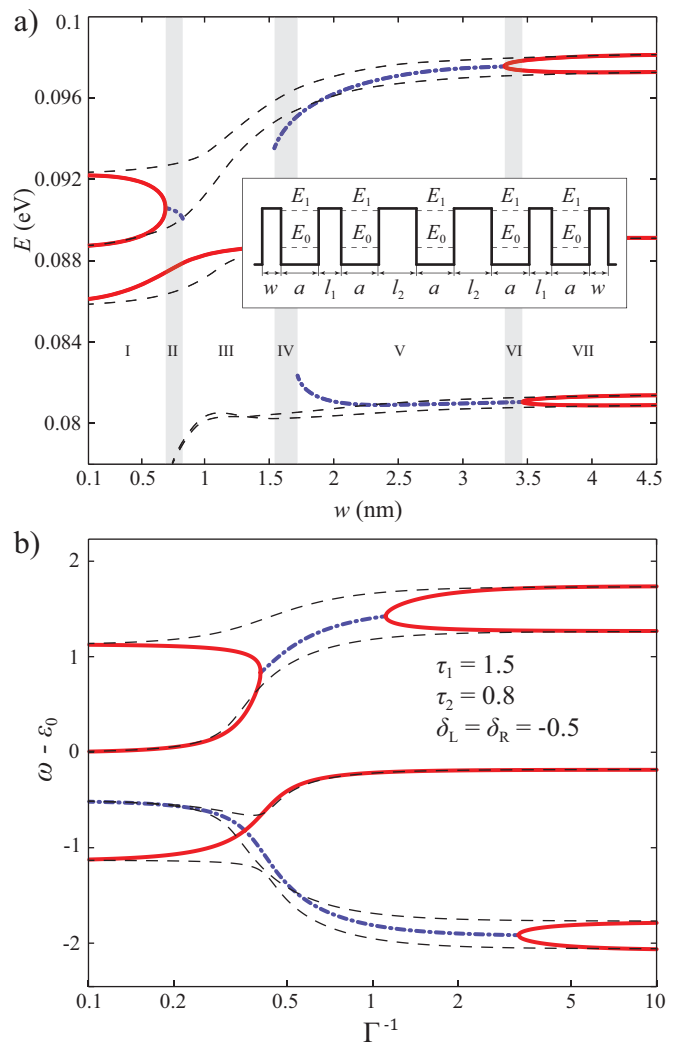}
\caption{\label{fig4}(Color online) Numerically calculated positions of the transmission maxima (solid lines -- unity values, dot-dashed lines -- non unity values) and the real parts of the S-matrix poles (dashed lines) vs. the outside barrier width $w$ for the structure (shown in the inset) with $N=5$ wells. Regions with different number of the transmission peaks are labeled by Roman numerals. The postcollapse regime with only one transmission peak is observed in region III. (a). The positions of the real eigenvalues of tight-binding auxiliary Hamiltonian (solid line) and the real parts of its complex eigenvalues (dot-dashed lines) along with the real parts of effective Hamiltonian eigenvalues (thin dashed lines) vs. the inverse rate of the decay into the contacts $\Gamma$ in the logarithmic scale for the structure with $N=5$ wells and the non-zero negative energy shifts $\delta_{L}=\delta_{R}<0$ in the outer wells (b).}
\end{figure}

It directly follows from the above presented physical picture of SB that in odd-$N$ structures a coalescence of two adjacent resonances will result in SB just as in even-$N$ structures. Consider a symmetric $N=5$ wells structure with the barrier height of $V=0.3$ eV, inner barriers thickness $l_{1}=3$ nm and $l_{2}=5$ nm (inset in the Fig.~\ref{fig4}a.) and interbarrier separation (well width) $a=5$ nm. In this case there will be two groups of levels of the whole structure (strictly speaking -- subbands accounting for the in-plane motion) originating from the two energy levels $E_{0}=0.0893$ eV and $E_{1}=0.2869$ eV in the isolated well. We consider the lowest group of levels, generated by the state with the energy $E_{0}<V/2$, which means that there are negative shifts $\delta_{L}=\delta_{R}$ in the 1-st and 5-th wells energies\cite{bib:Cel}, respectively. The results of the envelope function~(\ref{eqShr}) simulations are depicted in Fig.~\ref{fig4}a. Figure~\ref{fig4}b presents a plot of the real eigenvalues and real parts of the complex eigenvalues of the tight-binding auxiliary Hamiltonian that qualitatively corresponds to this situation ($\delta_{L}=\delta_{R}<0$). The tight-binding qualitative model helps to understand which pair of the resonances undergoes a coalescence and bifurcation because it shows the positions of the real parts of the complex eigenvalues of $H_{aux}$ in the region, where the numerical calculation fails to resolve nonunity peaks, as they are absorbed by peaks with higher transparency. The simulation results explicitly demonstrate that two adjacent resonances in an odd-$N$ structure do undergo CR accompanied by SB. Another interesting feature -- is a reentrant transition into a $\mathcal{PT}$-symmetric region from a symmetry broken phase. Analogous behavior of the eigenvalues was described in Ref.\cite{bib:IRot2,bib:Reentrant}.

A coalescence of only one pair of resonances in an $N=5$ well symmetric structure was described in Ref.\cite{bib:Cel}, where it was related to a superradiance transition and population trapping\cite{bib:Trap} caused by multichannel effects (the latter results in narrowing of $N-2$ and broadening of $2$ resonance levels of an open system\cite{bib:Mies}). Our structure (inset in Fig.~\ref{fig4}a) differs from the structure in Ref.\cite{bib:Cel} only by different numerical values of the barrier thickness, height and wells width. In our example one can observe a coalescence of different numbers of resonance pairs in different regions: three unity peaks in region I, one unity and one nonunity peak in region II, single unity peak in region III and so on (see Fig.~\ref{fig4}a). Hence it has nothing to do with multichannel effects (in fact we have only one incoming and one outgoing channel).

\section{Catastrophes}
\label{Sec.Cat}
Consider an $N$-order EP where $N$ resonances with the unit transparency coalesce (CR point). Such a point is defined by a certain ratio between the structure's parameters $\Gamma:|\tau_{1}|:...:|\tau_{N-1}|$, i.e. it is a specific point in the projective space of the parameters. Thus, one can treat an $N$-order EP as a quantum catastrophe\cite{bib:Zno2012,bib:LaZno2014} of the auxiliary Hamiltonian. At the very moment of the resonance coalescence the polynomial $Q$ has only one degenerate root $\omega=\varepsilon_{0}$ and so it has the form: $Q_{CR}=(\omega-\varepsilon_{0})^{N}$. Consequently, the transmission coefficient at the CR point has essentially a non Breit-Wigner form:
\begin{equation}
T_{NW}^{CR}(\omega)=\frac{\tilde\Gamma^{2N}}{(\omega-\varepsilon_{0})^{2N}+\tilde\Gamma^{2N}},
\label{eqTNcol}
\end{equation}
where $\tilde\Gamma$ is a function of structure's parameters $\Gamma$ and ${\tau_{i}}$.

The main feature of this expression is that first $2N-1$ derivatives take zero values at $\omega=\varepsilon_{0}$. The first nonzero derivative will be only of the $2N$-th order, which means that it is a degenerate extremum point, and the best mathematical tool here is the catastrophe theory. If one chooses the structures parameters $\{\tau_{i}\}$ and $\Gamma$ close to their critical ratios $|\tau_{i}|/\Gamma=c_{i}$ (corresponding to CR), the Taylor expansion of the transmission coefficient near $\omega=\varepsilon_{0}$ will be as follows (truncating higher order terms):
\begin{equation}
\begin{split}
f_{NW}^{0}(x)=&\frac{1}{a}\left[T_{NW}(\omega)-T_{NW}(\varepsilon_{0})\right]\approx\\
&x^{2N}+\sum_{k=2}^{2N-1}\frac{x^{k}}{ak!}T_{NW}^{(k)},
\label{eqEx}
\end{split}
\end{equation}
where $a=\frac{1}{(2N)!}T_{NW}^{(2N)}(\varepsilon_{0})\neq0$, $x=\omega-\varepsilon_{0}$ and $T'_{NW}(\varepsilon_{0})=0$ as it is an extremum point. If one introduces a small detuning of the parameters from their critical ratios: $\alpha_{i}=|\tau_{i}|-c_{i}\Gamma$, then higher order derivatives (up to $2N-1$) can be assumed to be some $h$-order ($h\geq 1$) forms of $\{\alpha_{i}\}$ vector, because they vanish at the moment of CR (when all $\alpha_{i}=0$).

According to the Tom and Arnold's classification of elementary catastrophes\cite{bib:Gilm} such an expansion with the highest nonzero term of the $2N$-th order is classified as of cuspoid type catastrophe $A_{-(2N-1)}$. Consider as an example expansions for triple- and four-barrier ($N=2$ and $N=3$) structures for which the catastrophe classification has special names: $A_{-3}$ -- cusp and $A_{-5}$ -- butterfly. After bringing to the canonical form (i.e. applying an appropriate variable substitution) the expansions take the form of cross-sections of the corresponding catastrophes:
\begin{equation}
\begin{split}
f_{2W}^{0}(x)=&x^4+k_{2}\alpha_{1}x^2,\\
f_{3W}^{0}(x)=&x^6+l_{4}\alpha_{1}x^4+l_{2}\alpha_{1}^{2}x^2,
\label{eqCB}
\end{split}
\end{equation}
where $k_{2}$ and $l_{2,4}$ are nonzero coefficients. Here $\alpha_{1}$ is a control parameter and $x$ is a variable.

\begin{figure}
\includegraphics{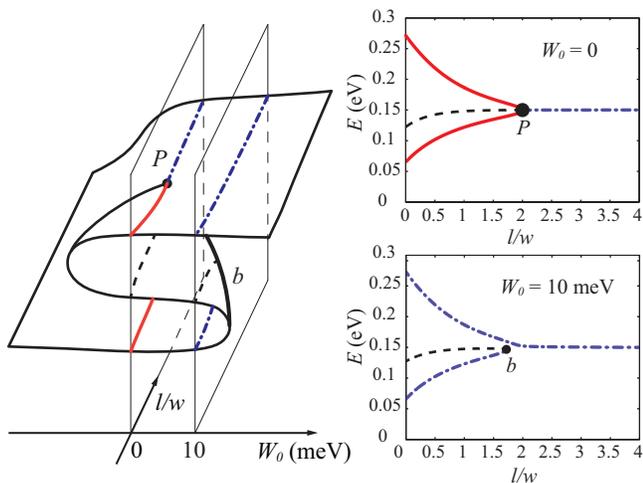}
\caption{\label{fig5}(Color online) On the left: schematic view of the extreme points surface of the total transmission of a symmetric double-well RTS in the neighborhood of CR point P. The horizontal axis are parameters (the central to the outside barriers widths ratio $l/w$ and the scattering $W_0$). The point $P$ -- is a CR point (EP), in which the maximums (upper and lower lists of the surface) and minimum (middle list of the surface) coalesce. On the right: numerically calculated cross-sections (in the planes shown on the left) of this surface for the structure without scattering (upper) and with $W_{0}=10$ meV (lower). The solid lines -- unity maximums of the transmission, the dot-dashed lines -- non unity maximums and the dashed lines -- minimums.}
\end{figure}

Introduction of small imperfections in an ideal symmetric structure results in a modification of the transmission coefficient derivatives in Eq.~(\ref{eqEx}). Dissipative effects in the Keldysh formalism can be described by an appropriate self-energy\cite{bib:Car2} with a certain spectral function $J(\omega)$ of the scatterers (in the simplest case of an elastic phase breaking\cite{bib:RDat}: $J(\omega)\propto\delta(\omega)$). Also, one can treat scattering in a phenomenological model of an optical potential\cite{bib:Zoh} in the effective mass Shr\"odinger equation approach~(\ref{eqShr}). The scattering influences the transmission in two ways: through a modification of the coherent part and through an introduction of the incoherent part. Weak elastic phase breaking can be described phenomenologically by adding small terms in the expressions for the derivatives. Therefore, the canonical expansion $f_{3B}^{0}$ in eq.~(\ref{eqCB}) can be rewritten as:
\begin{equation}
f_{2W}^{Scatt.}(x)=x^4 + (k_{2}\alpha_{1}+m_{2}W_{0})x^2 +m_{1}W_{0}x,
\label{eqDiss}
\end{equation}
where $m_{1,2}$ -- nonzero coefficients and $W_0$ -- a quantity, describing the scattering (for example, a value of the optical potential or the imaginary part of the corresponding self-energy in the Keldysh formalism). This expansion is a fully parametrized cusp catastrophe. From eq.~(\ref{eqDiss}) it is clear that CR is possible only in the absence of scattering (when both linear and quadratic terms in~(\ref{eqDiss}) turn to zero). Figure~\ref{fig5} schematically shows the surface of the extreme points of the total transmission (sum of the coherent and incoherent parts) of a symmetric double-well structure with scattering in the wells in the small neighborhood of CR point. Also, one can see here its numerically calculated cross-sections for the structure with left and right barrier thickness of $w=2$ nm, barriers height $V=0.3$ eV, well width $a=3.0485$ nm and effective mass $m^{*}=0.067 m_{0}$. This surface have a typical form of a cusp type catastrophe. It can be shown that presence of scattering modifies $P$ in the denominator of Eq.~(\ref{eqTN}) to $P'>P$, thus, making it greater than $P$ in the numerator.

Introduction of asymmetry in a double-well RTS ($\Gamma_{L}\neq\Gamma_{R}$) has similar results. If asymmetry is small, then the expansion of the transmission in the canonical form is expressed as:
\begin{equation}
f_{2W}^{Asymm.}(x)=x^4+(k_{2}\alpha_{1}+n_{2}\gamma)x^2+n_{1}\gamma\alpha_{1} x,
\label{eqAsymm}
\end{equation}
where $\gamma$ is a small asymmetry parameter: $\gamma=\Gamma_{L}-\Gamma_{R}$ and $n_{1,2}$ are nonzero coefficients. Contrary to scattering, an asymmetry influences the linear term only in the second order of smallness (multiplied by another small parameter $\alpha_{1}$). It is worth mentioning that, as can be seen from Eq.~(\ref{eqDiss}\,--\,\ref{eqAsymm}), effects of an asymmetry and scattering cannot cancel each other, i.e. CR cannot be achieved in the structure with both small scattering and a small asymmetry, because their influence is of different order. 

\section{Summary}
We have shown that the problem of determining the positions of the transparency peaks in dissipationless RTS can be mapped exactly on the eigenvalue problem for an auxiliary non-Hermitian Hamiltonian. This auxiliary non-Hermitian Hamiltonian can be deduced directly from an effective optical-potential-like Hamiltonian, which describes decaying states of an open quantum system and whose eigenvalues determine the poles of the scattering matrix. Real eigenvalues of the auxiliary non-Hermitian Hamiltonian $\it {exactly}$ define the position of unit transparency peaks. Hence, we obtain the procedure of mapping generally non-symmetric RTS on a family of non-Hermitian Hamiltonians, which generally do not have $\mathcal{PT}$-symmetry but do possess real eigenvalues. What kind of symmetry, if any, is responsible for realness of eigenvalues is an interesting question to be answered in future studies.

In a spatially symmetric RTS the auxiliary non-Hermitian Hamiltonian is $\mathcal{PT}$-symmetric and its real eigenvalues can coalesce at EP. Hence, all the results obtained in the literature concerning EP and $\mathcal{PT}$-symmetry breaking are applicable to the description of a symmetric RTS. A coalescence of the auxiliary non-Hermitian Hamiltonian eigenvalues means the coalescence of resonances in RTS. Thus, manifestation of $\mathcal{PT}$-SB in a fermionic system has been described. In fermionic systems a conventional approach to associate $\mathcal{PT}$-invariant terms in the non-Hermitian Hamiltonian with gain/loss processes is not directly applicable because of the particle (norm) conserving requirement for fermions. In our model these terms correspond to incoming/outgoing flows of electrons. A macroscopic quantity, which characterizes $\mathcal{PT}$-symmetry of our system is the stationary current or particle flow that is a polar and time-odd vector. Our model explains why a coalescence of an even number of resonances does result in SB and a coalescence of an odd number of resonances does not. Spontaneous SB phenomenon in an open quantum system has been already described earlier\cite{bib:CalLeg} within the Caldeira-Legget model\cite{bib:CalLegOr}, which relates SSB to tunneling suppressed by dissipation. In our case SB is a purely coherent phenomenon. Dissipation destroys EP and SB. 

In condensed matter physics a classification of states with broken symmetry is based on the group theory, which limits the possible number of different states. In the case of CR in RTS the mirror symmetry is the only symmetry that is broken. However, the number of resonances which coalesce can vary. We have shown that a variety of exceptional points and related phenomena characterized by coalescence of a different number of resonances can be classified in terms of the catastrophe theory. 

A direct experimental observation of a coalescence of resonances in RTS is not a simple task. In common current-voltage characteristic measurements the peculiarities of the transparency coefficient near EP in the expression for the current could be smeared out by a contribution of electrons with different energies within the extended Fermi-distribution. Hence, delicate differential transport measurements and/or a implementation of selective in energy electron sources are required. The latter can be realized, for example, with the help of traditional technique based on resonant tunneling\cite{bib:Capasso}.  Another feature, which can have possible applications is a specific non Breit-Wigner transmission profile~(\ref{eqTNcol}). In optical systems, where, contrary to electron systems, monochromaticity is essential, transmission peculiarities can be observed directly. So, such a profile creates a wide  window with almost perfect transmission.

\begin{acknowledgments}
AAG would like to acknowledge the Program of Fundamental Research of the Presidium of the Russian Academy of Science for partial support. 
\end{acknowledgments}

\appendix

\section{Properties of polynomial $Q$ in symmetric case}
\label{ApA}
In the case of a symmetric structure ($\Gamma_{L}=\Gamma_{R}=\Gamma$ and $\tau_{i}=\tau_{N-i}$) with all the restrictions mentioned in Sec.~\ref{Sec.Symm} ($\varepsilon_{i}=\varepsilon_{0}$ and $\delta_{L}=\delta_{R}=0$) the polynomial $Q$ is the characteristic polynomial of the $\mathcal{PT}$-symmetric tridiagonal matrix $(\omega\hat I-\hat H_{\mathcal{PT}})$ with $\hat H_{\mathcal{PT}}$ from~(\ref{eqHPT}). So we can use appropriate relations and write $Q$ as follows:
\begin{multline}
Q=\det{\left(\omega\hat I-\hat H_{\mathcal{PT}}\right)}\\
=\left(\omega^{2}+\Gamma^{2}\right)D_{1}^{1}-2\omega|\tau_{1}|^{2}D_{1}^{2}+|\tau_{1}|^{4}D_{2}^{2}.
\label{eqApQ}
\end{multline}
Here and everywhere below in this section we assume $\varepsilon_{0}\mapsto0$ without loss of generality.

Consider the case of an even number of quantum wells $N$. The statement to prove here is that $Q$ is a polynomial of $\omega^{2}$, i.e. it consists only of even powers of $\omega$. This is obvious for $N=2$ as $Q_{2W}=\omega^{2}+\Gamma^{2}-|\tau_{1}|^{2}$. For an even $N>2$ we will, first of all, prove the following statement:
\begin{equation}
\begin{split}
\left(D_{1}^{1}\right)_{N}=&p_{N}(\omega^{2}),\\
\left(D_{1}^{2}\right)_{N}=&\omega q_{N}(\omega^{2}),\\
\left(D_{2}^{2}\right)_{N}=&r_{N}(\omega^{2}),
\label{eqApPr1}
\end{split}
\end{equation}
where $p_{N}$, $q_{N}$ and $r_{N}$ are some polynomials and the subscript $N$ indicates the number of wells, i.e. the dimensions of the initial matrix $(\omega\hat I-\hat H_{\mathcal{PT}})$. We use the mathematical induction on an even $N$. The induction basis is the $N=4$ case: $\left(D_{1}^{1}\right)_{4}=\omega^{2}-|\tau_{2}|^{2}$, $\left(D_{1}^{2}\right)_{4}=\omega$ and $\left(D_{2}^{2}\right)_{4}=1$. Now we assume that~(\ref{eqApPr1}) has been proved for $N-2$ and try to prove it for $N$. So, using the tridiagonal matrix properties one can expand $\left(D_{1}^{1}\right)_{N}$:
\begin{equation}
\left(D_{1}^{1}\right)_{N}=\omega^{2}\left(D_{2}^{2}\right)_{N}-2\omega|\tau_{2}|^{2}\left(D_{3}^{2}\right)_{N}+|\tau_{2}|^{4}\left(D_{3}^{3}\right)_{N}.
\label{eqAp3}
\end{equation}
It is easy to see that the quantity $(D_{i}^{j})_{N}$ is the same as $(D_{i-1}^{j-1})_{N-2}$ under some appropriate reindexing of $\tau_{i}$. Hence, according to induction assumption~(\ref{eqApPr1}) for $N-2$ we can write following:
\begin{equation}
\begin{split}
\left(D_{2}^{2}\right)_{N}=&\tilde p_{N-2}(\omega^{2}),\\
\left(D_{1}^{2}\right)_{N}=&\omega\tilde q_{N-2}(\omega^{2}),\\
\left(D_{2}^{2}\right)_{N}=&\tilde r_{N-2}(\omega^{2}),
\label{eqAp4}
\end{split}
\end{equation}
where $\tilde p_{N-2}$, $\tilde q_{N-2}$ and $\tilde r_{N-2}$ are the corresponding polynomials from~(\ref{eqApPr1}) with the reindexed $\tau_{i}$. Thus, substituting~(\ref{eqAp4}) into~(\ref{eqAp3}), we show that $\left(D_{1}^{1}\right)_{N}$ is a polynomial of $\omega^{2}$, which proves the first equality in~(\ref{eqApPr1}). The second and the third equalities in~(\ref{eqApPr1}) can be proved by the same arguments. So, after showing that~(\ref{eqApPr1}) is true, we can substitute it into~(\ref{eqApQ}) and get that for an even $N$ polynomial $Q$ is the $N/2$-th order polynomial of $\omega^{2}$.

For an odd $N$ in the same way one can prove that $Q$ is a polynomial of $\omega^{2}$ multiplied by $\omega$, i.e. it consists only of odd powers of $\omega$. Cases $N=1$ and $N=3$ are easily checked straightforwardly and for an odd $N>3$ once again we can use the mathematical induction just as it was done above for an even $N$. So, first we show that
\begin{equation}
\begin{split}
\left(D_{1}^{1}\right)_{N}=&\omega p_{N}(\omega^{2}),\\
\left(D_{1}^{2}\right)_{N}=&q_{N}(\omega^{2}),\\
\left(D_{2}^{2}\right)_{N}=&\omega r_{N}(\omega^{2}).
\label{eqApPr2}
\end{split}
\end{equation}
Then substituting~(\ref{eqApPr2}) into~(\ref{eqApQ}) finishes the proof of the polynomial $Q$ to be the $(N-1)/2$-th order polynomial of $\omega^{2}$ multiplied by $\omega$ in the case of an odd $N$.

\bibliography{references}

\begin{thebibliography}{67}%
\makeatletter
\providecommand \@ifxundefined [1]{%
 \@ifx{#1\undefined}
}%
\providecommand \@ifnum [1]{%
 \ifnum #1\expandafter \@firstoftwo
 \else \expandafter \@secondoftwo
 \fi
}%
\providecommand \@ifx [1]{%
 \ifx #1\expandafter \@firstoftwo
 \else \expandafter \@secondoftwo
 \fi
}%
\providecommand \natexlab [1]{#1}%
\providecommand \enquote  [1]{``#1''}%
\providecommand \bibnamefont  [1]{#1}%
\providecommand \bibfnamefont [1]{#1}%
\providecommand \citenamefont [1]{#1}%
\providecommand \href@noop [0]{\@secondoftwo}%
\providecommand \href [0]{\begingroup \@sanitize@url \@href}%
\providecommand \@href[1]{\@@startlink{#1}\@@href}%
\providecommand \@@href[1]{\endgroup#1\@@endlink}%
\providecommand \@sanitize@url [0]{\catcode `\\12\catcode `\$12\catcode
  `\&12\catcode `\#12\catcode `\^12\catcode `\_12\catcode `\%12\relax}%
\providecommand \@@startlink[1]{}%
\providecommand \@@endlink[0]{}%
\providecommand \url  [0]{\begingroup\@sanitize@url \@url }%
\providecommand \@url [1]{\endgroup\@href {#1}{\urlprefix }}%
\providecommand \urlprefix  [0]{URL }%
\providecommand \Eprint [0]{\href }%
\providecommand \doibase [0]{http://dx.doi.org/}%
\providecommand \selectlanguage [0]{\@gobble}%
\providecommand \bibinfo  [0]{\@secondoftwo}%
\providecommand \bibfield  [0]{\@secondoftwo}%
\providecommand \translation [1]{[#1]}%
\providecommand \BibitemOpen [0]{}%
\providecommand \bibitemStop [0]{}%
\providecommand \bibitemNoStop [0]{.\EOS\space}%
\providecommand \EOS [0]{\spacefactor3000\relax}%
\providecommand \BibitemShut  [1]{\csname bibitem#1\endcsname}%
\let\auto@bib@innerbib\@empty
\bibitem [{\citenamefont {Bender}\ and\ \citenamefont
  {Boettcher}(1998)}]{bib:PTBend98}%
  \BibitemOpen
  \bibfield  {author} {\bibinfo {author} {\bibfnamefont {C.~M.}\ \bibnamefont
  {Bender}}\ and\ \bibinfo {author} {\bibfnamefont {S.}~\bibnamefont
  {Boettcher}},\ }\href {\doibase 10.1103/PhysRevLett.80.5243} {\bibfield
  {journal} {\bibinfo  {journal} {Phys. Rev. Lett.}\ }\textbf {\bibinfo
  {volume} {80}},\ \bibinfo {pages} {5243} (\bibinfo {year}
  {1998})}\BibitemShut {NoStop}%
\bibitem [{\citenamefont {Bender}\ \emph {et~al.}(1999)\citenamefont {Bender},
  \citenamefont {Boettcher},\ and\ \citenamefont {Meisinger}}]{bib:PTBend99}%
  \BibitemOpen
  \bibfield  {author} {\bibinfo {author} {\bibfnamefont {C.~M.}\ \bibnamefont
  {Bender}}, \bibinfo {author} {\bibfnamefont {S.}~\bibnamefont {Boettcher}}, \
  and\ \bibinfo {author} {\bibfnamefont {P.~N.}\ \bibnamefont {Meisinger}},\
  }\href {\doibase http://dx.doi.org/10.1063/1.532860} {\bibfield  {journal}
  {\bibinfo  {journal} {Journal of Mathematical Physics}\ }\textbf {\bibinfo
  {volume} {40}},\ \bibinfo {pages} {2201} (\bibinfo {year}
  {1999})}\BibitemShut {NoStop}%
\bibitem [{\citenamefont {Bender}(2007)}]{bib:PTBend07}%
  \BibitemOpen
  \bibfield  {author} {\bibinfo {author} {\bibfnamefont {C.~M.}\ \bibnamefont
  {Bender}},\ }\href {http://stacks.iop.org/0034-4885/70/i=6/a=R03} {\bibfield
  {journal} {\bibinfo  {journal} {Reports on Progress in Physics}\ }\textbf
  {\bibinfo {volume} {70}},\ \bibinfo {pages} {947} (\bibinfo {year}
  {2007})}\BibitemShut {NoStop}%
\bibitem [{\citenamefont {Mostafazadeh}(2002)}]{bib:mostafa2002}%
  \BibitemOpen
  \bibfield  {author} {\bibinfo {author} {\bibfnamefont {A.}~\bibnamefont
  {Mostafazadeh}},\ }\href@noop {} {\bibfield  {journal} {\bibinfo  {journal}
  {J. Math. Phys.}\ }\textbf {\bibinfo {volume} {43}} (\bibinfo {year}
  {2002})}\BibitemShut {NoStop}%
\bibitem [{\citenamefont {Kato}(1995)}]{bib:BookKato}%
  \BibitemOpen
  \bibfield  {author} {\bibinfo {author} {\bibfnamefont {T.}~\bibnamefont
  {Kato}},\ }\href {https://books.google.ru/books?id=8ji2kN\_D3BwC} {\emph
  {\bibinfo {title} {Perturbation Theory for Linear Operators}}},\ Classics in
  Mathematics\ (\bibinfo  {publisher} {Springer Berlin Heidelberg},\ \bibinfo
  {year} {1995})\BibitemShut {NoStop}%
\bibitem [{\citenamefont {Heiss}(2012)}]{bib:Heiss}%
  \BibitemOpen
  \bibfield  {author} {\bibinfo {author} {\bibfnamefont {W.~D.}\ \bibnamefont
  {Heiss}},\ }\href {http://stacks.iop.org/1751-8121/45/i=44/a=444016}
  {\bibfield  {journal} {\bibinfo  {journal} {Journal of Physics A:
  Mathematical and Theoretical}\ }\textbf {\bibinfo {volume} {45}},\ \bibinfo
  {pages} {444016} (\bibinfo {year} {2012})}\BibitemShut {NoStop}%
\bibitem [{\citenamefont {Berry}(2004)}]{bib:Berry}%
  \BibitemOpen
  \bibfield  {author} {\bibinfo {author} {\bibfnamefont {M.}~\bibnamefont
  {Berry}},\ }\href@noop {} {\bibfield  {journal} {\bibinfo  {journal}
  {Czechoslovak journal of physics}\ }\textbf {\bibinfo {volume} {54}},\
  \bibinfo {pages} {1039} (\bibinfo {year} {2004})}\BibitemShut {NoStop}%
\bibitem [{\citenamefont {Wiersig}(2016)}]{bib:EPWier}%
  \BibitemOpen
  \bibfield  {author} {\bibinfo {author} {\bibfnamefont {J.}~\bibnamefont
  {Wiersig}},\ }\href {\doibase 10.1103/PhysRevA.93.033809} {\bibfield
  {journal} {\bibinfo  {journal} {Phys. Rev. A}\ }\textbf {\bibinfo {volume}
  {93}},\ \bibinfo {pages} {033809} (\bibinfo {year} {2016})}\BibitemShut
  {NoStop}%
\bibitem [{\citenamefont {Feng}\ \emph {et~al.}(2014)\citenamefont {Feng},
  \citenamefont {Zhu}, \citenamefont {Yang}, \citenamefont {Zhu}, \citenamefont
  {Zhang}, \citenamefont {Yin}, \citenamefont {Wang},\ and\ \citenamefont
  {Zhang}}]{bib:NonRec}%
  \BibitemOpen
  \bibfield  {author} {\bibinfo {author} {\bibfnamefont {L.}~\bibnamefont
  {Feng}}, \bibinfo {author} {\bibfnamefont {X.}~\bibnamefont {Zhu}}, \bibinfo
  {author} {\bibfnamefont {S.}~\bibnamefont {Yang}}, \bibinfo {author}
  {\bibfnamefont {H.}~\bibnamefont {Zhu}}, \bibinfo {author} {\bibfnamefont
  {P.}~\bibnamefont {Zhang}}, \bibinfo {author} {\bibfnamefont
  {X.}~\bibnamefont {Yin}}, \bibinfo {author} {\bibfnamefont {Y.}~\bibnamefont
  {Wang}}, \ and\ \bibinfo {author} {\bibfnamefont {X.}~\bibnamefont {Zhang}},\
  }\href {\doibase 10.1364/OE.22.001760} {\bibfield  {journal} {\bibinfo
  {journal} {Opt. Express}\ }\textbf {\bibinfo {volume} {22}},\ \bibinfo
  {pages} {1760} (\bibinfo {year} {2014})}\BibitemShut {NoStop}%
\bibitem [{\citenamefont {Feng}\ \emph {et~al.}(2011)\citenamefont {Feng},
  \citenamefont {Ayache}, \citenamefont {Huang}, \citenamefont {Xu},
  \citenamefont {Lu}, \citenamefont {Chen}, \citenamefont {Fainman},\ and\
  \citenamefont {Scherer}}]{bib:PTopt2}%
  \BibitemOpen
  \bibfield  {author} {\bibinfo {author} {\bibfnamefont {L.}~\bibnamefont
  {Feng}}, \bibinfo {author} {\bibfnamefont {M.}~\bibnamefont {Ayache}},
  \bibinfo {author} {\bibfnamefont {J.}~\bibnamefont {Huang}}, \bibinfo
  {author} {\bibfnamefont {Y.-L.}\ \bibnamefont {Xu}}, \bibinfo {author}
  {\bibfnamefont {M.-H.}\ \bibnamefont {Lu}}, \bibinfo {author} {\bibfnamefont
  {Y.-F.}\ \bibnamefont {Chen}}, \bibinfo {author} {\bibfnamefont
  {Y.}~\bibnamefont {Fainman}}, \ and\ \bibinfo {author} {\bibfnamefont
  {A.}~\bibnamefont {Scherer}},\ }\href {\doibase 10.1126/science.1206038}
  {\bibfield  {journal} {\bibinfo  {journal} {Science}\ }\textbf {\bibinfo
  {volume} {333}},\ \bibinfo {pages} {729} (\bibinfo {year}
  {2011})}\BibitemShut {NoStop}%
\bibitem [{\citenamefont {Doppler}\ \emph {et~al.}(2016)\citenamefont
  {Doppler}, \citenamefont {Mailybaev}, \citenamefont {B{\"o}hm}, \citenamefont
  {Kuhl}, \citenamefont {Girschik}, \citenamefont {Libisch}, \citenamefont
  {Milburn}, \citenamefont {Rabl}, \citenamefont {Moiseyev},\ and\
  \citenamefont {Rotter}}]{bib:Switch}%
  \BibitemOpen
  \bibfield  {author} {\bibinfo {author} {\bibfnamefont {J.}~\bibnamefont
  {Doppler}}, \bibinfo {author} {\bibfnamefont {A.~A.}\ \bibnamefont
  {Mailybaev}}, \bibinfo {author} {\bibfnamefont {J.}~\bibnamefont {B{\"o}hm}},
  \bibinfo {author} {\bibfnamefont {U.}~\bibnamefont {Kuhl}}, \bibinfo {author}
  {\bibfnamefont {A.}~\bibnamefont {Girschik}}, \bibinfo {author}
  {\bibfnamefont {F.}~\bibnamefont {Libisch}}, \bibinfo {author} {\bibfnamefont
  {T.~J.}\ \bibnamefont {Milburn}}, \bibinfo {author} {\bibfnamefont
  {P.}~\bibnamefont {Rabl}}, \bibinfo {author} {\bibfnamefont {N.}~\bibnamefont
  {Moiseyev}}, \ and\ \bibinfo {author} {\bibfnamefont {S.}~\bibnamefont
  {Rotter}},\ }\href {http://dx.doi.org/10.1038/nature18605} {\bibfield
  {journal} {\bibinfo  {journal} {Nature}\ }\textbf {\bibinfo {volume} {advance
  online publication}},\  (\bibinfo {year} {2016})}\BibitemShut {NoStop}%
\bibitem [{\citenamefont {Zyablovsky}\ \emph {et~al.}(2014)\citenamefont
  {Zyablovsky}, \citenamefont {Vinogradov}, \citenamefont {Pukhov},
  \citenamefont {Dorofeenko},\ and\ \citenamefont {Lisyansky}}]{bib:Zyab}%
  \BibitemOpen
  \bibfield  {author} {\bibinfo {author} {\bibfnamefont {A.~A.}\ \bibnamefont
  {Zyablovsky}}, \bibinfo {author} {\bibfnamefont {A.~P.}\ \bibnamefont
  {Vinogradov}}, \bibinfo {author} {\bibfnamefont {A.~A.}\ \bibnamefont
  {Pukhov}}, \bibinfo {author} {\bibfnamefont {A.~V.}\ \bibnamefont
  {Dorofeenko}}, \ and\ \bibinfo {author} {\bibfnamefont {A.~A.}\ \bibnamefont
  {Lisyansky}},\ }\href {\doibase 10.3367/UFNe.0184.201411b.1177} {\bibfield
  {journal} {\bibinfo  {journal} {Physics-Uspekhi}\ }\textbf {\bibinfo {volume}
  {57}},\ \bibinfo {pages} {1063} (\bibinfo {year} {2014})}\BibitemShut
  {NoStop}%
\bibitem [{\citenamefont {Guo}\ \emph {et~al.}(2009)\citenamefont {Guo},
  \citenamefont {Salamo}, \citenamefont {Duchesne}, \citenamefont {Morandotti},
  \citenamefont {Volatier-Ravat}, \citenamefont {Aimez}, \citenamefont
  {Siviloglou},\ and\ \citenamefont {Christodoulides}}]{bib:PTopt1}%
  \BibitemOpen
  \bibfield  {author} {\bibinfo {author} {\bibfnamefont {A.}~\bibnamefont
  {Guo}}, \bibinfo {author} {\bibfnamefont {G.~J.}\ \bibnamefont {Salamo}},
  \bibinfo {author} {\bibfnamefont {D.}~\bibnamefont {Duchesne}}, \bibinfo
  {author} {\bibfnamefont {R.}~\bibnamefont {Morandotti}}, \bibinfo {author}
  {\bibfnamefont {M.}~\bibnamefont {Volatier-Ravat}}, \bibinfo {author}
  {\bibfnamefont {V.}~\bibnamefont {Aimez}}, \bibinfo {author} {\bibfnamefont
  {G.~A.}\ \bibnamefont {Siviloglou}}, \ and\ \bibinfo {author} {\bibfnamefont
  {D.~N.}\ \bibnamefont {Christodoulides}},\ }\href {\doibase
  10.1103/PhysRevLett.103.093902} {\bibfield  {journal} {\bibinfo  {journal}
  {Phys. Rev. Lett.}\ }\textbf {\bibinfo {volume} {103}},\ \bibinfo {pages}
  {093902} (\bibinfo {year} {2009})}\BibitemShut {NoStop}%
\bibitem [{\citenamefont {Regensburger}\ \emph {et~al.}(2012)\citenamefont
  {Regensburger}, \citenamefont {Bersch}, \citenamefont {Miri}, \citenamefont
  {Onishchukov}, \citenamefont {Christodoulides},\ and\ \citenamefont
  {Peschel}}]{bib:NatPTLight}%
  \BibitemOpen
  \bibfield  {author} {\bibinfo {author} {\bibfnamefont {A.}~\bibnamefont
  {Regensburger}}, \bibinfo {author} {\bibfnamefont {C.}~\bibnamefont
  {Bersch}}, \bibinfo {author} {\bibfnamefont {M.-A.}\ \bibnamefont {Miri}},
  \bibinfo {author} {\bibfnamefont {G.}~\bibnamefont {Onishchukov}}, \bibinfo
  {author} {\bibfnamefont {D.~N.}\ \bibnamefont {Christodoulides}}, \ and\
  \bibinfo {author} {\bibfnamefont {U.}~\bibnamefont {Peschel}},\ }\href
  {\doibase 10.1038/nature11298} {\bibfield  {journal} {\bibinfo  {journal}
  {Nature}\ }\textbf {\bibinfo {volume} {488}},\ \bibinfo {pages} {167}
  (\bibinfo {year} {2012})}\BibitemShut {NoStop}%
\bibitem [{\citenamefont {Liertzer}\ \emph {et~al.}(2012)\citenamefont
  {Liertzer}, \citenamefont {Ge}, \citenamefont {Cerjan}, \citenamefont
  {Stone}, \citenamefont {T\"ureci},\ and\ \citenamefont {Rotter}}]{bib:Laser}%
  \BibitemOpen
  \bibfield  {author} {\bibinfo {author} {\bibfnamefont {M.}~\bibnamefont
  {Liertzer}}, \bibinfo {author} {\bibfnamefont {L.}~\bibnamefont {Ge}},
  \bibinfo {author} {\bibfnamefont {A.}~\bibnamefont {Cerjan}}, \bibinfo
  {author} {\bibfnamefont {A.~D.}\ \bibnamefont {Stone}}, \bibinfo {author}
  {\bibfnamefont {H.~E.}\ \bibnamefont {T\"ureci}}, \ and\ \bibinfo {author}
  {\bibfnamefont {S.}~\bibnamefont {Rotter}},\ }\href {\doibase
  10.1103/PhysRevLett.108.173901} {\bibfield  {journal} {\bibinfo  {journal}
  {Phys. Rev. Lett.}\ }\textbf {\bibinfo {volume} {108}},\ \bibinfo {pages}
  {173901} (\bibinfo {year} {2012})}\BibitemShut {NoStop}%
\bibitem [{\citenamefont {Brandstetter}\ \emph {et~al.}(2014)\citenamefont
  {Brandstetter}, \citenamefont {Liertzer}, \citenamefont {Deutsch},
  \citenamefont {Klang}, \citenamefont {Sch{\"o}berl}, \citenamefont
  {T{\"u}reci}, \citenamefont {Strasser}, \citenamefont {Unterrainer},\ and\
  \citenamefont {Rotter}}]{bib:nature}%
  \BibitemOpen
  \bibfield  {author} {\bibinfo {author} {\bibfnamefont {M.}~\bibnamefont
  {Brandstetter}}, \bibinfo {author} {\bibfnamefont {M.}~\bibnamefont
  {Liertzer}}, \bibinfo {author} {\bibfnamefont {C.}~\bibnamefont {Deutsch}},
  \bibinfo {author} {\bibfnamefont {P.}~\bibnamefont {Klang}}, \bibinfo
  {author} {\bibfnamefont {J.}~\bibnamefont {Sch{\"o}berl}}, \bibinfo {author}
  {\bibfnamefont {H.~E.}\ \bibnamefont {T{\"u}reci}}, \bibinfo {author}
  {\bibfnamefont {G.}~\bibnamefont {Strasser}}, \bibinfo {author}
  {\bibfnamefont {K.}~\bibnamefont {Unterrainer}}, \ and\ \bibinfo {author}
  {\bibfnamefont {S.}~\bibnamefont {Rotter}},\ }\href
  {http://dx.doi.org/10.1038/ncomms5034} {\bibfield  {journal} {\bibinfo
  {journal} {Nat Commun}\ }\textbf {\bibinfo {volume} {5}},\ \bibinfo {pages}
  {4034} (\bibinfo {year} {2014})}\BibitemShut {NoStop}%
\bibitem [{\citenamefont {Chong}\ \emph {et~al.}(2010)\citenamefont {Chong},
  \citenamefont {Ge}, \citenamefont {Cao},\ and\ \citenamefont
  {Stone}}]{bib:CPACho}%
  \BibitemOpen
  \bibfield  {author} {\bibinfo {author} {\bibfnamefont {Y.~D.}\ \bibnamefont
  {Chong}}, \bibinfo {author} {\bibfnamefont {L.}~\bibnamefont {Ge}}, \bibinfo
  {author} {\bibfnamefont {H.}~\bibnamefont {Cao}}, \ and\ \bibinfo {author}
  {\bibfnamefont {A.~D.}\ \bibnamefont {Stone}},\ }\href {\doibase
  10.1103/PhysRevLett.105.053901} {\bibfield  {journal} {\bibinfo  {journal}
  {Phys. Rev. Lett.}\ }\textbf {\bibinfo {volume} {105}},\ \bibinfo {pages}
  {053901} (\bibinfo {year} {2010})}\BibitemShut {NoStop}%
\bibitem [{\citenamefont {Longhi}(2010)}]{bib:CPALon}%
  \BibitemOpen
  \bibfield  {author} {\bibinfo {author} {\bibfnamefont {S.}~\bibnamefont
  {Longhi}},\ }\href {\doibase 10.1103/PhysRevA.82.031801} {\bibfield
  {journal} {\bibinfo  {journal} {Phys. Rev. A}\ }\textbf {\bibinfo {volume}
  {82}},\ \bibinfo {pages} {031801} (\bibinfo {year} {2010})}\BibitemShut
  {NoStop}%
\bibitem [{\citenamefont {Chong}\ \emph {et~al.}(2011)\citenamefont {Chong},
  \citenamefont {Ge},\ and\ \citenamefont {Stone}}]{bib:Cho}%
  \BibitemOpen
  \bibfield  {author} {\bibinfo {author} {\bibfnamefont {Y.~D.}\ \bibnamefont
  {Chong}}, \bibinfo {author} {\bibfnamefont {L.}~\bibnamefont {Ge}}, \ and\
  \bibinfo {author} {\bibfnamefont {A.~D.}\ \bibnamefont {Stone}},\ }\href
  {\doibase 10.1103/PhysRevLett.106.093902} {\bibfield  {journal} {\bibinfo
  {journal} {Phys. Rev. Lett.}\ }\textbf {\bibinfo {volume} {106}},\ \bibinfo
  {pages} {093902} (\bibinfo {year} {2011})}\BibitemShut {NoStop}%
\bibitem [{\citenamefont {Ambichl}\ \emph {et~al.}(2013)\citenamefont
  {Ambichl}, \citenamefont {Makris}, \citenamefont {Ge}, \citenamefont {Chong},
  \citenamefont {Stone},\ and\ \citenamefont {Rotter}}]{bib:Ambi}%
  \BibitemOpen
  \bibfield  {author} {\bibinfo {author} {\bibfnamefont {P.}~\bibnamefont
  {Ambichl}}, \bibinfo {author} {\bibfnamefont {K.~G.}\ \bibnamefont {Makris}},
  \bibinfo {author} {\bibfnamefont {L.}~\bibnamefont {Ge}}, \bibinfo {author}
  {\bibfnamefont {Y.}~\bibnamefont {Chong}}, \bibinfo {author} {\bibfnamefont
  {A.~D.}\ \bibnamefont {Stone}}, \ and\ \bibinfo {author} {\bibfnamefont
  {S.}~\bibnamefont {Rotter}},\ }\href {\doibase 10.1103/PhysRevX.3.041030}
  {\bibfield  {journal} {\bibinfo  {journal} {Phys. Rev. X}\ }\textbf {\bibinfo
  {volume} {3}},\ \bibinfo {pages} {041030} (\bibinfo {year}
  {2013})}\BibitemShut {NoStop}%
\bibitem [{\citenamefont {Bittner}\ \emph {et~al.}(2012)\citenamefont
  {Bittner}, \citenamefont {Dietz}, \citenamefont {G\"unther}, \citenamefont
  {Harney}, \citenamefont {Miski-Oglu}, \citenamefont {Richter},\ and\
  \citenamefont {Sch\"afer}}]{bib:PTmicroW}%
  \BibitemOpen
  \bibfield  {author} {\bibinfo {author} {\bibfnamefont {S.}~\bibnamefont
  {Bittner}}, \bibinfo {author} {\bibfnamefont {B.}~\bibnamefont {Dietz}},
  \bibinfo {author} {\bibfnamefont {U.}~\bibnamefont {G\"unther}}, \bibinfo
  {author} {\bibfnamefont {H.~L.}\ \bibnamefont {Harney}}, \bibinfo {author}
  {\bibfnamefont {M.}~\bibnamefont {Miski-Oglu}}, \bibinfo {author}
  {\bibfnamefont {A.}~\bibnamefont {Richter}}, \ and\ \bibinfo {author}
  {\bibfnamefont {F.}~\bibnamefont {Sch\"afer}},\ }\href {\doibase
  10.1103/PhysRevLett.108.024101} {\bibfield  {journal} {\bibinfo  {journal}
  {Phys. Rev. Lett.}\ }\textbf {\bibinfo {volume} {108}},\ \bibinfo {pages}
  {024101} (\bibinfo {year} {2012})}\BibitemShut {NoStop}%
\bibitem [{\citenamefont {Feng}(2016)}]{bib:Feng2016}%
  \BibitemOpen
  \bibfield  {author} {\bibinfo {author} {\bibfnamefont {S.}~\bibnamefont
  {Feng}},\ }\href {\doibase 10.1364/OE.24.001291} {\bibfield  {journal}
  {\bibinfo  {journal} {Opt. Express}\ }\textbf {\bibinfo {volume} {24}},\
  \bibinfo {pages} {1291} (\bibinfo {year} {2016})}\BibitemShut {NoStop}%
\bibitem [{\citenamefont {Kreibich}\ \emph {et~al.}(2013)\citenamefont
  {Kreibich}, \citenamefont {Main}, \citenamefont {Cartarius},\ and\
  \citenamefont {Wunner}}]{bib:BEC}%
  \BibitemOpen
  \bibfield  {author} {\bibinfo {author} {\bibfnamefont {M.}~\bibnamefont
  {Kreibich}}, \bibinfo {author} {\bibfnamefont {J.}~\bibnamefont {Main}},
  \bibinfo {author} {\bibfnamefont {H.}~\bibnamefont {Cartarius}}, \ and\
  \bibinfo {author} {\bibfnamefont {G.}~\bibnamefont {Wunner}},\ }\href
  {\doibase 10.1103/PhysRevA.87.051601} {\bibfield  {journal} {\bibinfo
  {journal} {Phys. Rev. A}\ }\textbf {\bibinfo {volume} {87}},\ \bibinfo
  {pages} {051601} (\bibinfo {year} {2013})}\BibitemShut {NoStop}%
\bibitem [{\citenamefont {Chtchelkatchev}\ \emph {et~al.}(2012)\citenamefont
  {Chtchelkatchev}, \citenamefont {Golubov}, \citenamefont {Baturina},\ and\
  \citenamefont {Vinokur}}]{bib:PTsc}%
  \BibitemOpen
  \bibfield  {author} {\bibinfo {author} {\bibfnamefont {N.~M.}\ \bibnamefont
  {Chtchelkatchev}}, \bibinfo {author} {\bibfnamefont {A.~A.}\ \bibnamefont
  {Golubov}}, \bibinfo {author} {\bibfnamefont {T.~I.}\ \bibnamefont
  {Baturina}}, \ and\ \bibinfo {author} {\bibfnamefont {V.~M.}\ \bibnamefont
  {Vinokur}},\ }\href {\doibase 10.1103/PhysRevLett.109.150405} {\bibfield
  {journal} {\bibinfo  {journal} {Phys. Rev. Lett.}\ }\textbf {\bibinfo
  {volume} {109}},\ \bibinfo {pages} {150405} (\bibinfo {year}
  {2012})}\BibitemShut {NoStop}%
\bibitem [{\citenamefont {San-Jose}\ \emph {et~al.}(2016)\citenamefont
  {San-Jose}, \citenamefont {Cayao}, \citenamefont {Prada},\ and\ \citenamefont
  {Aguado}}]{bib:Madrid}%
  \BibitemOpen
  \bibfield  {author} {\bibinfo {author} {\bibfnamefont {P.}~\bibnamefont
  {San-Jose}}, \bibinfo {author} {\bibfnamefont {J.}~\bibnamefont {Cayao}},
  \bibinfo {author} {\bibfnamefont {E.}~\bibnamefont {Prada}}, \ and\ \bibinfo
  {author} {\bibfnamefont {R.}~\bibnamefont {Aguado}},\ }\href@noop {}
  {\bibfield  {journal} {\bibinfo  {journal} {Scientific Reports}\ }\textbf
  {\bibinfo {volume} {6}},\ \bibinfo {pages} {21427 EP } (\bibinfo {year}
  {2016})},\ \bibinfo {note} {article}\BibitemShut {NoStop}%
\bibitem [{\citenamefont {Mandal}(2015)}]{bib:M44}%
  \BibitemOpen
  \bibfield  {author} {\bibinfo {author} {\bibfnamefont {I.}~\bibnamefont
  {Mandal}},\ }\href@noop {} {\bibfield  {journal} {\bibinfo  {journal} {EPL
  (Europhysics Letters)}\ }\textbf {\bibinfo {volume} {110}},\ \bibinfo {pages}
  {67005} (\bibinfo {year} {2015})}\BibitemShut {NoStop}%
\bibitem [{\citenamefont {Newton}(2013)}]{bib:Newton}%
  \BibitemOpen
  \bibfield  {author} {\bibinfo {author} {\bibfnamefont {R.}~\bibnamefont
  {Newton}},\ }\href {https://books.google.ru/books?id=KWTyCAAAQBAJ} {\emph
  {\bibinfo {title} {Scattering Theory of Waves and Particles}}},\ Theoretical
  and Mathematical Physics\ (\bibinfo  {publisher} {Springer Berlin
  Heidelberg},\ \bibinfo {year} {2013})\BibitemShut {NoStop}%
\bibitem [{\citenamefont {Moiseyev}(2011)}]{bib:BookMo}%
  \BibitemOpen
  \bibfield  {author} {\bibinfo {author} {\bibfnamefont {N.}~\bibnamefont
  {Moiseyev}},\ }\href@noop {} {\emph {\bibinfo {title} {Non-Hermitian quantum
  mechanics}}}\ (\bibinfo  {publisher} {Cambridge University Press},\ \bibinfo
  {year} {2011})\BibitemShut {NoStop}%
\bibitem [{\citenamefont {Grecchi}\ and\ \citenamefont
  {Sacchetti}(1995)}]{bib:grecchi}%
  \BibitemOpen
  \bibfield  {author} {\bibinfo {author} {\bibfnamefont {V.}~\bibnamefont
  {Grecchi}}\ and\ \bibinfo {author} {\bibfnamefont {A.}~\bibnamefont
  {Sacchetti}},\ }\href {\doibase http://dx.doi.org/10.1006/aphy.1995.1063}
  {\bibfield  {journal} {\bibinfo  {journal} {Annals of Physics}\ }\textbf
  {\bibinfo {volume} {241}},\ \bibinfo {pages} {258 } (\bibinfo {year}
  {1995})}\BibitemShut {NoStop}%
\bibitem [{\citenamefont {Gorbatsevich}\ \emph {et~al.}(2008)\citenamefont
  {Gorbatsevich}, \citenamefont {Zhuravlev},\ and\ \citenamefont
  {Kapaev}}]{bib:Gor}%
  \BibitemOpen
  \bibfield  {author} {\bibinfo {author} {\bibfnamefont {A.}~\bibnamefont
  {Gorbatsevich}}, \bibinfo {author} {\bibfnamefont {M.}~\bibnamefont
  {Zhuravlev}}, \ and\ \bibinfo {author} {\bibfnamefont {V.}~\bibnamefont
  {Kapaev}},\ }\href {\doibase 10.1134/S106377610808013X} {\bibfield  {journal}
  {\bibinfo  {journal} {Journal of Experimental and Theoretical Physics}\
  }\textbf {\bibinfo {volume} {107}},\ \bibinfo {pages} {288} (\bibinfo {year}
  {2008})}\BibitemShut {NoStop}%
\bibitem [{\citenamefont {Nussenzveig}(1959)}]{bib:Spole1}%
  \BibitemOpen
  \bibfield  {author} {\bibinfo {author} {\bibfnamefont {H.}~\bibnamefont
  {Nussenzveig}},\ }\href {\doibase
  http://dx.doi.org/10.1016/0029-5582(59)90293-7} {\bibfield  {journal}
  {\bibinfo  {journal} {Nuclear Physics}\ }\textbf {\bibinfo {volume} {11}},\
  \bibinfo {pages} {499 } (\bibinfo {year} {1959})}\BibitemShut {NoStop}%
\bibitem [{\citenamefont {Klaiman}\ and\ \citenamefont
  {Moiseyev}(2010)}]{bib:AbsEn}%
  \BibitemOpen
  \bibfield  {author} {\bibinfo {author} {\bibfnamefont {S.}~\bibnamefont
  {Klaiman}}\ and\ \bibinfo {author} {\bibfnamefont {N.}~\bibnamefont
  {Moiseyev}},\ }\href {http://stacks.iop.org/0953-4075/43/i=18/a=185205}
  {\bibfield  {journal} {\bibinfo  {journal} {Journal of Physics B: Atomic,
  Molecular and Optical Physics}\ }\textbf {\bibinfo {volume} {43}},\ \bibinfo
  {pages} {185205} (\bibinfo {year} {2010})}\BibitemShut {NoStop}%
\bibitem [{\citenamefont {Tolstikhin}\ \emph {et~al.}(1997)\citenamefont
  {Tolstikhin}, \citenamefont {Ostrovsky},\ and\ \citenamefont
  {Nakamura}}]{bib:Tolstikhin}%
  \BibitemOpen
  \bibfield  {author} {\bibinfo {author} {\bibfnamefont {O.~I.}\ \bibnamefont
  {Tolstikhin}}, \bibinfo {author} {\bibfnamefont {V.~N.}\ \bibnamefont
  {Ostrovsky}}, \ and\ \bibinfo {author} {\bibfnamefont {H.}~\bibnamefont
  {Nakamura}},\ }\href {\doibase 10.1103/PhysRevLett.79.2026} {\bibfield
  {journal} {\bibinfo  {journal} {Phys. Rev. Lett.}\ }\textbf {\bibinfo
  {volume} {79}},\ \bibinfo {pages} {2026} (\bibinfo {year}
  {1997})}\BibitemShut {NoStop}%
\bibitem [{\citenamefont {M\"uller}\ \emph {et~al.}(1995)\citenamefont
  {M\"uller}, \citenamefont {Dittes}, \citenamefont {Iskra},\ and\
  \citenamefont {Rotter}}]{bib:rotter1995}%
  \BibitemOpen
  \bibfield  {author} {\bibinfo {author} {\bibfnamefont {M.}~\bibnamefont
  {M\"uller}}, \bibinfo {author} {\bibfnamefont {F.-M.}\ \bibnamefont
  {Dittes}}, \bibinfo {author} {\bibfnamefont {W.}~\bibnamefont {Iskra}}, \
  and\ \bibinfo {author} {\bibfnamefont {I.}~\bibnamefont {Rotter}},\ }\href
  {\doibase 10.1103/PhysRevE.52.5961} {\bibfield  {journal} {\bibinfo
  {journal} {Phys. Rev. E}\ }\textbf {\bibinfo {volume} {52}},\ \bibinfo
  {pages} {5961} (\bibinfo {year} {1995})}\BibitemShut {NoStop}%
\bibitem [{\citenamefont {Garmon}\ \emph {et~al.}(2015)\citenamefont {Garmon},
  \citenamefont {Gianfreda},\ and\ \citenamefont {Hatano}}]{bib:Res}%
  \BibitemOpen
  \bibfield  {author} {\bibinfo {author} {\bibfnamefont {S.}~\bibnamefont
  {Garmon}}, \bibinfo {author} {\bibfnamefont {M.}~\bibnamefont {Gianfreda}}, \
  and\ \bibinfo {author} {\bibfnamefont {N.}~\bibnamefont {Hatano}},\ }\href
  {\doibase 10.1103/PhysRevA.92.022125} {\bibfield  {journal} {\bibinfo
  {journal} {Phys. Rev. A}\ }\textbf {\bibinfo {volume} {92}},\ \bibinfo
  {pages} {022125} (\bibinfo {year} {2015})}\BibitemShut {NoStop}%
\bibitem [{\citenamefont {Rotter}\ and\ \citenamefont
  {Bird}(2015)}]{bib:rotter2015}%
  \BibitemOpen
  \bibfield  {author} {\bibinfo {author} {\bibfnamefont {I.}~\bibnamefont
  {Rotter}}\ and\ \bibinfo {author} {\bibfnamefont {J.~P.}\ \bibnamefont
  {Bird}},\ }\href {http://stacks.iop.org/0034-4885/78/i=11/a=114001}
  {\bibfield  {journal} {\bibinfo  {journal} {Reports on Progress in Physics}\
  }\textbf {\bibinfo {volume} {78}},\ \bibinfo {pages} {114001} (\bibinfo
  {year} {2015})}\BibitemShut {NoStop}%
\bibitem [{\citenamefont {Ge}\ \emph {et~al.}(2012)\citenamefont {Ge},
  \citenamefont {Chong},\ and\ \citenamefont {Stone}}]{bib:SmatrEig}%
  \BibitemOpen
  \bibfield  {author} {\bibinfo {author} {\bibfnamefont {L.}~\bibnamefont
  {Ge}}, \bibinfo {author} {\bibfnamefont {Y.~D.}\ \bibnamefont {Chong}}, \
  and\ \bibinfo {author} {\bibfnamefont {A.~D.}\ \bibnamefont {Stone}},\ }\href
  {\doibase 10.1103/PhysRevA.85.023802} {\bibfield  {journal} {\bibinfo
  {journal} {Phys. Rev. A}\ }\textbf {\bibinfo {volume} {85}},\ \bibinfo
  {pages} {023802} (\bibinfo {year} {2012})}\BibitemShut {NoStop}%
\bibitem [{\citenamefont {Hernandez-Coronado}\ \emph
  {et~al.}(2011)\citenamefont {Hernandez-Coronado}, \citenamefont
  {Krej\v{c}i\v{r}\'{i}k},\ and\ \citenamefont {Siegl}}]{bib:PTScat}%
  \BibitemOpen
  \bibfield  {author} {\bibinfo {author} {\bibfnamefont {H.}~\bibnamefont
  {Hernandez-Coronado}}, \bibinfo {author} {\bibfnamefont {D.}~\bibnamefont
  {Krej\v{c}i\v{r}\'{i}k}}, \ and\ \bibinfo {author} {\bibfnamefont
  {P.}~\bibnamefont {Siegl}},\ }\href {\doibase
  http://dx.doi.org/10.1016/j.physleta.2011.04.021} {\bibfield  {journal}
  {\bibinfo  {journal} {Physics Letters A}\ }\textbf {\bibinfo {volume}
  {375}},\ \bibinfo {pages} {2149 } (\bibinfo {year} {2011})}\BibitemShut
  {NoStop}%
\bibitem [{\citenamefont {Gorbatsevich}\ and\ \citenamefont
  {Shubin}(2016)}]{bib:GorSh}%
  \BibitemOpen
  \bibfield  {author} {\bibinfo {author} {\bibfnamefont {A.~A.}\ \bibnamefont
  {Gorbatsevich}}\ and\ \bibinfo {author} {\bibfnamefont {N.~M.}\ \bibnamefont
  {Shubin}},\ }\href {\doibase 10.1134/S0021364016120031} {\bibfield  {journal}
  {\bibinfo  {journal} {JETP Letters}\ ,\ \bibinfo {pages} {1}} (\bibinfo
  {year} {2016})}\BibitemShut {NoStop}%
\bibitem [{\citenamefont {Gilmore}(1992)}]{bib:Gilm}%
  \BibitemOpen
  \bibfield  {author} {\bibinfo {author} {\bibfnamefont {R.}~\bibnamefont
  {Gilmore}},\ }\href@noop {} {\bibfield  {journal} {\bibinfo  {journal}
  {Encyclopedia of applied physics}\ }\textbf {\bibinfo {volume} {3}},\
  \bibinfo {pages} {85} (\bibinfo {year} {1992})}\BibitemShut {NoStop}%
\bibitem [{\citenamefont {Rotter}\ and\ \citenamefont
  {Sadreev}(2004)}]{bib:IRot2004}%
  \BibitemOpen
  \bibfield  {author} {\bibinfo {author} {\bibfnamefont {I.}~\bibnamefont
  {Rotter}}\ and\ \bibinfo {author} {\bibfnamefont {A.~F.}\ \bibnamefont
  {Sadreev}},\ }\href {\doibase 10.1103/PhysRevE.69.066201} {\bibfield
  {journal} {\bibinfo  {journal} {Phys. Rev. E}\ }\textbf {\bibinfo {volume}
  {69}},\ \bibinfo {pages} {066201} (\bibinfo {year} {2004})}\BibitemShut
  {NoStop}%
\bibitem [{\citenamefont {Caroli}\ \emph {et~al.}(1971)\citenamefont {Caroli},
  \citenamefont {Combescot}, \citenamefont {Nozieres},\ and\ \citenamefont
  {Saint-James}}]{bib:Car1}%
  \BibitemOpen
  \bibfield  {author} {\bibinfo {author} {\bibfnamefont {C.}~\bibnamefont
  {Caroli}}, \bibinfo {author} {\bibfnamefont {R.}~\bibnamefont {Combescot}},
  \bibinfo {author} {\bibfnamefont {P.}~\bibnamefont {Nozieres}}, \ and\
  \bibinfo {author} {\bibfnamefont {D.}~\bibnamefont {Saint-James}},\ }\href
  {http://stacks.iop.org/0022-3719/4/i=8/a=018} {\bibfield  {journal} {\bibinfo
   {journal} {Journal of Physics C: Solid State Physics}\ }\textbf {\bibinfo
  {volume} {4}},\ \bibinfo {pages} {916} (\bibinfo {year} {1971})}\BibitemShut
  {NoStop}%
\bibitem [{\citenamefont {Kopaev}\ and\ \citenamefont
  {Molotkov}(1994)}]{bib:Kapa}%
  \BibitemOpen
  \bibfield  {author} {\bibinfo {author} {\bibfnamefont {Y.~V.}\ \bibnamefont
  {Kopaev}}\ and\ \bibinfo {author} {\bibfnamefont {S.~N.}\ \bibnamefont
  {Molotkov}},\ }\href@noop {} {\bibfield  {journal} {\bibinfo  {journal} {JETP
  Letters}\ }\textbf {\bibinfo {volume} {59}},\ \bibinfo {pages} {800}
  (\bibinfo {year} {1994})}\BibitemShut {NoStop}%
\bibitem [{\citenamefont {Feshbach}(1958)}]{bib:F1}%
  \BibitemOpen
  \bibfield  {author} {\bibinfo {author} {\bibfnamefont {H.}~\bibnamefont
  {Feshbach}},\ }\href {\doibase
  http://dx.doi.org/10.1016/0003-4916(58)90007-1} {\bibfield  {journal}
  {\bibinfo  {journal} {Annals of Physics}\ }\textbf {\bibinfo {volume} {5}},\
  \bibinfo {pages} {357 } (\bibinfo {year} {1958})}\BibitemShut {NoStop}%
\bibitem [{\citenamefont {Feshbach}(1962)}]{bib:F2}%
  \BibitemOpen
  \bibfield  {author} {\bibinfo {author} {\bibfnamefont {H.}~\bibnamefont
  {Feshbach}},\ }\href {\doibase
  http://dx.doi.org/10.1016/0003-4916(62)90221-X} {\bibfield  {journal}
  {\bibinfo  {journal} {Annals of Physics}\ }\textbf {\bibinfo {volume} {19}},\
  \bibinfo {pages} {287 } (\bibinfo {year} {1962})}\BibitemShut {NoStop}%
\bibitem [{\citenamefont {Celardo}\ and\ \citenamefont
  {Kaplan}(2009)}]{bib:Cel}%
  \BibitemOpen
  \bibfield  {author} {\bibinfo {author} {\bibfnamefont {G.~L.}\ \bibnamefont
  {Celardo}}\ and\ \bibinfo {author} {\bibfnamefont {L.}~\bibnamefont
  {Kaplan}},\ }\href {\doibase 10.1103/PhysRevB.79.155108} {\bibfield
  {journal} {\bibinfo  {journal} {Phys. Rev. B}\ }\textbf {\bibinfo {volume}
  {79}},\ \bibinfo {pages} {155108} (\bibinfo {year} {2009})}\BibitemShut
  {NoStop}%
\bibitem [{\citenamefont {Celardo}\ \emph {et~al.}(2010)\citenamefont
  {Celardo}, \citenamefont {Smith}, \citenamefont {Sorathia}, \citenamefont
  {Zelevinsky}, \citenamefont {Sen'kov},\ and\ \citenamefont
  {Kaplan}}]{bib:Celardo2010}%
  \BibitemOpen
  \bibfield  {author} {\bibinfo {author} {\bibfnamefont {G.~L.}\ \bibnamefont
  {Celardo}}, \bibinfo {author} {\bibfnamefont {A.~M.}\ \bibnamefont {Smith}},
  \bibinfo {author} {\bibfnamefont {S.}~\bibnamefont {Sorathia}}, \bibinfo
  {author} {\bibfnamefont {V.~G.}\ \bibnamefont {Zelevinsky}}, \bibinfo
  {author} {\bibfnamefont {R.~A.}\ \bibnamefont {Sen'kov}}, \ and\ \bibinfo
  {author} {\bibfnamefont {L.}~\bibnamefont {Kaplan}},\ }\href {\doibase
  10.1103/PhysRevB.82.165437} {\bibfield  {journal} {\bibinfo  {journal} {Phys.
  Rev. B}\ }\textbf {\bibinfo {volume} {82}},\ \bibinfo {pages} {165437}
  (\bibinfo {year} {2010})}\BibitemShut {NoStop}%
\bibitem [{\citenamefont {Jin}\ and\ \citenamefont {Song}(2010)}]{bib:Jin2010}%
  \BibitemOpen
  \bibfield  {author} {\bibinfo {author} {\bibfnamefont {L.}~\bibnamefont
  {Jin}}\ and\ \bibinfo {author} {\bibfnamefont {Z.}~\bibnamefont {Song}},\
  }\href {\doibase 10.1103/PhysRevA.81.032109} {\bibfield  {journal} {\bibinfo
  {journal} {Phys. Rev. A}\ }\textbf {\bibinfo {volume} {81}},\ \bibinfo
  {pages} {032109} (\bibinfo {year} {2010})}\BibitemShut {NoStop}%
\bibitem [{\citenamefont {Dente}\ \emph {et~al.}(2008)\citenamefont {Dente},
  \citenamefont {Bustos-Mar\'un},\ and\ \citenamefont
  {Pastawski}}]{bib:RealSE2}%
  \BibitemOpen
  \bibfield  {author} {\bibinfo {author} {\bibfnamefont {A.~D.}\ \bibnamefont
  {Dente}}, \bibinfo {author} {\bibfnamefont {R.~A.}\ \bibnamefont
  {Bustos-Mar\'un}}, \ and\ \bibinfo {author} {\bibfnamefont {H.~M.}\
  \bibnamefont {Pastawski}},\ }\href {\doibase 10.1103/PhysRevA.78.062116}
  {\bibfield  {journal} {\bibinfo  {journal} {Phys. Rev. A}\ }\textbf {\bibinfo
  {volume} {78}},\ \bibinfo {pages} {062116} (\bibinfo {year}
  {2008})}\BibitemShut {NoStop}%
\bibitem [{\citenamefont {Ryndyk}\ \emph {et~al.}(2009)\citenamefont {Ryndyk},
  \citenamefont {Guti{\'e}rrez}, \citenamefont {Song},\ and\ \citenamefont
  {Cuniberti}}]{bib:RealSE1}%
  \BibitemOpen
  \bibfield  {author} {\bibinfo {author} {\bibfnamefont {D.}~\bibnamefont
  {Ryndyk}}, \bibinfo {author} {\bibfnamefont {R.}~\bibnamefont
  {Guti{\'e}rrez}}, \bibinfo {author} {\bibfnamefont {B.}~\bibnamefont {Song}},
  \ and\ \bibinfo {author} {\bibfnamefont {G.}~\bibnamefont {Cuniberti}},\ }in\
  \href@noop {} {\emph {\bibinfo {booktitle} {Energy Transfer Dynamics in
  Biomaterial Systems}}}\ (\bibinfo  {publisher} {Springer},\ \bibinfo {year}
  {2009})\ pp.\ \bibinfo {pages} {213--335}\BibitemShut {NoStop}%
\bibitem [{\citenamefont {Eleuch}\ and\ \citenamefont
  {Rotter}(2015)}]{bib:IRot1}%
  \BibitemOpen
  \bibfield  {author} {\bibinfo {author} {\bibfnamefont {H.}~\bibnamefont
  {Eleuch}}\ and\ \bibinfo {author} {\bibfnamefont {I.}~\bibnamefont
  {Rotter}},\ }\href {\doibase 10.1140/epjd/e2015-60390-2} {\bibfield
  {journal} {\bibinfo  {journal} {The European Physical Journal D}\ }\textbf
  {\bibinfo {volume} {69}},\ \bibinfo {pages} {1} (\bibinfo {year}
  {2015})}\BibitemShut {NoStop}%
\bibitem [{\citenamefont {Eleuch}\ and\ \citenamefont
  {Rotter}(2016)}]{bib:IRot2}%
  \BibitemOpen
  \bibfield  {author} {\bibinfo {author} {\bibfnamefont {H.}~\bibnamefont
  {Eleuch}}\ and\ \bibinfo {author} {\bibfnamefont {I.}~\bibnamefont
  {Rotter}},\ }\href {\doibase 10.1103/PhysRevA.93.042116} {\bibfield
  {journal} {\bibinfo  {journal} {Phys. Rev. A}\ }\textbf {\bibinfo {volume}
  {93}},\ \bibinfo {pages} {042116} (\bibinfo {year} {2016})}\BibitemShut
  {NoStop}%
\bibitem [{\citenamefont {Dhar}\ and\ \citenamefont {Sen}(2006)}]{bib:ElDen}%
  \BibitemOpen
  \bibfield  {author} {\bibinfo {author} {\bibfnamefont {A.}~\bibnamefont
  {Dhar}}\ and\ \bibinfo {author} {\bibfnamefont {D.}~\bibnamefont {Sen}},\
  }\href {\doibase 10.1103/PhysRevB.73.085119} {\bibfield  {journal} {\bibinfo
  {journal} {Phys. Rev. B}\ }\textbf {\bibinfo {volume} {73}},\ \bibinfo
  {pages} {085119} (\bibinfo {year} {2006})}\BibitemShut {NoStop}%
\bibitem [{\citenamefont {Sheng}\ and\ \citenamefont {Xia}(1996)}]{bib:TM}%
  \BibitemOpen
  \bibfield  {author} {\bibinfo {author} {\bibfnamefont {W.-D.}\ \bibnamefont
  {Sheng}}\ and\ \bibinfo {author} {\bibfnamefont {J.-B.}\ \bibnamefont
  {Xia}},\ }\href {http://stacks.iop.org/0953-8984/8/i=20/a=009} {\bibfield
  {journal} {\bibinfo  {journal} {Journal of Physics: Condensed Matter}\
  }\textbf {\bibinfo {volume} {8}},\ \bibinfo {pages} {3635} (\bibinfo {year}
  {1996})}\BibitemShut {NoStop}%
\bibitem [{\citenamefont {Razavy}(2003)}]{bib:bookTM}%
  \BibitemOpen
  \bibfield  {author} {\bibinfo {author} {\bibfnamefont {M.}~\bibnamefont
  {Razavy}},\ }\href@noop {} {\emph {\bibinfo {title} {Quantum theory of
  tunneling}}}\ (\bibinfo  {publisher} {World Scientific},\ \bibinfo {year}
  {2003})\BibitemShut {NoStop}%
\bibitem [{\citenamefont {Loran}\ and\ \citenamefont
  {Mostafazadeh}(2016)}]{bib:TMallDim}%
  \BibitemOpen
  \bibfield  {author} {\bibinfo {author} {\bibfnamefont {F.}~\bibnamefont
  {Loran}}\ and\ \bibinfo {author} {\bibfnamefont {A.}~\bibnamefont
  {Mostafazadeh}},\ }\href {\doibase 10.1103/PhysRevA.93.042707} {\bibfield
  {journal} {\bibinfo  {journal} {Phys. Rev. A}\ }\textbf {\bibinfo {volume}
  {93}},\ \bibinfo {pages} {042707} (\bibinfo {year} {2016})}\BibitemShut
  {NoStop}%
\bibitem [{\citenamefont {Ding}\ \emph {et~al.}(2016)\citenamefont {Ding},
  \citenamefont {Ma}, \citenamefont {Xiao}, \citenamefont {Zhang},\ and\
  \citenamefont {Chan}}]{bib:Reentrant}%
  \BibitemOpen
  \bibfield  {author} {\bibinfo {author} {\bibfnamefont {K.}~\bibnamefont
  {Ding}}, \bibinfo {author} {\bibfnamefont {G.}~\bibnamefont {Ma}}, \bibinfo
  {author} {\bibfnamefont {M.}~\bibnamefont {Xiao}}, \bibinfo {author}
  {\bibfnamefont {Z.~Q.}\ \bibnamefont {Zhang}}, \ and\ \bibinfo {author}
  {\bibfnamefont {C.~T.}\ \bibnamefont {Chan}},\ }\href {\doibase
  10.1103/PhysRevX.6.021007} {\bibfield  {journal} {\bibinfo  {journal} {Phys.
  Rev. X}\ }\textbf {\bibinfo {volume} {6}},\ \bibinfo {pages} {021007}
  (\bibinfo {year} {2016})}\BibitemShut {NoStop}%
\bibitem [{\citenamefont {Persson}\ and\ \citenamefont
  {Rotter}(1999)}]{bib:Trap}%
  \BibitemOpen
  \bibfield  {author} {\bibinfo {author} {\bibfnamefont {E.}~\bibnamefont
  {Persson}}\ and\ \bibinfo {author} {\bibfnamefont {I.}~\bibnamefont
  {Rotter}},\ }\href {\doibase 10.1103/PhysRevC.59.164} {\bibfield  {journal}
  {\bibinfo  {journal} {Phys. Rev. C}\ }\textbf {\bibinfo {volume} {59}},\
  \bibinfo {pages} {164} (\bibinfo {year} {1999})}\BibitemShut {NoStop}%
\bibitem [{\citenamefont {Mies}(1968)}]{bib:Mies}%
  \BibitemOpen
  \bibfield  {author} {\bibinfo {author} {\bibfnamefont {F.~H.}\ \bibnamefont
  {Mies}},\ }\href {\doibase 10.1103/PhysRev.175.164} {\bibfield  {journal}
  {\bibinfo  {journal} {Phys. Rev.}\ }\textbf {\bibinfo {volume} {175}},\
  \bibinfo {pages} {164} (\bibinfo {year} {1968})}\BibitemShut {NoStop}%
\bibitem [{\citenamefont {Znojil}(2012)}]{bib:Zno2012}%
  \BibitemOpen
  \bibfield  {author} {\bibinfo {author} {\bibfnamefont {M.}~\bibnamefont
  {Znojil}},\ }\href@noop {} {\bibfield  {journal} {\bibinfo  {journal}
  {Journal of Physics A: Mathematical and Theoretical}\ }\textbf {\bibinfo
  {volume} {45}},\ \bibinfo {pages} {444036} (\bibinfo {year}
  {2012})}\BibitemShut {NoStop}%
\bibitem [{\citenamefont {L{\'e}vai}\ \emph {et~al.}(2014)\citenamefont
  {L{\'e}vai}, \citenamefont {R{\r{u}}{\v{z}}i{\v{c}}ka},\ and\ \citenamefont
  {Znojil}}]{bib:LaZno2014}%
  \BibitemOpen
  \bibfield  {author} {\bibinfo {author} {\bibfnamefont {G.}~\bibnamefont
  {L{\'e}vai}}, \bibinfo {author} {\bibfnamefont {F.}~\bibnamefont
  {R{\r{u}}{\v{z}}i{\v{c}}ka}}, \ and\ \bibinfo {author} {\bibfnamefont
  {M.}~\bibnamefont {Znojil}},\ }\href {\doibase 10.1007/s10773-014-2085-x}
  {\bibfield  {journal} {\bibinfo  {journal} {International Journal of
  Theoretical Physics}\ }\textbf {\bibinfo {volume} {53}},\ \bibinfo {pages}
  {2875} (\bibinfo {year} {2014})}\BibitemShut {NoStop}%
\bibitem [{\citenamefont {Caroli}\ \emph {et~al.}(1972)\citenamefont {Caroli},
  \citenamefont {Combescot}, \citenamefont {Nozieres},\ and\ \citenamefont
  {Saint-James}}]{bib:Car2}%
  \BibitemOpen
  \bibfield  {author} {\bibinfo {author} {\bibfnamefont {C.}~\bibnamefont
  {Caroli}}, \bibinfo {author} {\bibfnamefont {R.}~\bibnamefont {Combescot}},
  \bibinfo {author} {\bibfnamefont {P.}~\bibnamefont {Nozieres}}, \ and\
  \bibinfo {author} {\bibfnamefont {D.}~\bibnamefont {Saint-James}},\ }\href
  {http://stacks.iop.org/0022-3719/5/i=1/a=006} {\bibfield  {journal} {\bibinfo
   {journal} {Journal of Physics C: Solid State Physics}\ }\textbf {\bibinfo
  {volume} {5}},\ \bibinfo {pages} {21} (\bibinfo {year} {1972})}\BibitemShut
  {NoStop}%
\bibitem [{\citenamefont {Lake}\ and\ \citenamefont {Datta}(1992)}]{bib:RDat}%
  \BibitemOpen
  \bibfield  {author} {\bibinfo {author} {\bibfnamefont {R.}~\bibnamefont
  {Lake}}\ and\ \bibinfo {author} {\bibfnamefont {S.}~\bibnamefont {Datta}},\
  }\href {\doibase 10.1103/PhysRevB.45.6670} {\bibfield  {journal} {\bibinfo
  {journal} {Phys. Rev. B}\ }\textbf {\bibinfo {volume} {45}},\ \bibinfo
  {pages} {6670} (\bibinfo {year} {1992})}\BibitemShut {NoStop}%
\bibitem [{\citenamefont {Zohta}\ and\ \citenamefont {Ezawa}(1992)}]{bib:Zoh}%
  \BibitemOpen
  \bibfield  {author} {\bibinfo {author} {\bibfnamefont {Y.}~\bibnamefont
  {Zohta}}\ and\ \bibinfo {author} {\bibfnamefont {H.}~\bibnamefont {Ezawa}},\
  }\href {\doibase http://dx.doi.org/10.1063/1.352296} {\bibfield  {journal}
  {\bibinfo  {journal} {Journal of Applied Physics}\ }\textbf {\bibinfo
  {volume} {72}},\ \bibinfo {pages} {3584} (\bibinfo {year}
  {1992})}\BibitemShut {NoStop}%
\bibitem [{\citenamefont {Bray}\ and\ \citenamefont
  {Moore}(1982)}]{bib:CalLeg}%
  \BibitemOpen
  \bibfield  {author} {\bibinfo {author} {\bibfnamefont {A.~J.}\ \bibnamefont
  {Bray}}\ and\ \bibinfo {author} {\bibfnamefont {M.~A.}\ \bibnamefont
  {Moore}},\ }\href {\doibase 10.1103/PhysRevLett.49.1545} {\bibfield
  {journal} {\bibinfo  {journal} {Phys. Rev. Lett.}\ }\textbf {\bibinfo
  {volume} {49}},\ \bibinfo {pages} {1545} (\bibinfo {year}
  {1982})}\BibitemShut {NoStop}%
\bibitem [{\citenamefont {Caldeira}\ and\ \citenamefont
  {Leggett}(1981)}]{bib:CalLegOr}%
  \BibitemOpen
  \bibfield  {author} {\bibinfo {author} {\bibfnamefont {A.~O.}\ \bibnamefont
  {Caldeira}}\ and\ \bibinfo {author} {\bibfnamefont {A.~J.}\ \bibnamefont
  {Leggett}},\ }\href {\doibase 10.1103/PhysRevLett.46.211} {\bibfield
  {journal} {\bibinfo  {journal} {Phys. Rev. Lett.}\ }\textbf {\bibinfo
  {volume} {46}},\ \bibinfo {pages} {211} (\bibinfo {year} {1981})}\BibitemShut
  {NoStop}%
\bibitem [{\citenamefont {Capasso}\ \emph {et~al.}(1986)\citenamefont
  {Capasso}, \citenamefont {Mohammed},\ and\ \citenamefont
  {Cho}}]{bib:Capasso}%
  \BibitemOpen
  \bibfield  {author} {\bibinfo {author} {\bibfnamefont {F.}~\bibnamefont
  {Capasso}}, \bibinfo {author} {\bibfnamefont {K.}~\bibnamefont {Mohammed}}, \
  and\ \bibinfo {author} {\bibfnamefont {A.}~\bibnamefont {Cho}},\ }\href
  {\doibase 10.1109/JQE.1986.1073171} {\bibfield  {journal} {\bibinfo
  {journal} {IEEE Journal of Quantum Electronics}\ }\textbf {\bibinfo {volume}
  {22}},\ \bibinfo {pages} {1853} (\bibinfo {year} {1986})}\BibitemShut
  {NoStop}%
\end{thebibliography}%

\end{document}